\documentclass[lettersize,journal]{IEEEtran}
\usepackage{amsmath,amsfonts}
\usepackage{algorithm}
\usepackage{array}
\usepackage[caption=false,font=normalsize,labelfont=sf,textfont=sf]{subfig}
\usepackage{textcomp}
\usepackage{stfloats}
\usepackage{url}
\usepackage{verbatim}
\usepackage{graphicx}
\usepackage{cite}
\usepackage{multirow}
\usepackage{booktabs,tabularx}
\usepackage{tikz}
\usepackage{pifont}
\usepackage{algpseudocode}
\usepackage{amssymb}
\usepackage{orcidlink}
\usepackage{bm}
\usepackage{subcaption} %

\AtBeginDocument{%
  \setlength{\textfloatsep}{5pt plus 2pt minus 2pt}%
  \setlength{\floatsep}{6pt plus 2pt minus 2pt}%
  \setlength{\intextsep}{6pt plus 2pt minus 2pt}%
  \setlength{\abovecaptionskip}{3pt}%
  \setlength{\belowcaptionskip}{1pt}%
}
\usetikzlibrary{shapes.geometric, arrows.meta, positioning, calc, fit}

\usepackage{caption}
\begin{document}

\title{Learning to Look Benign: Targeted Evasion of Malware Detectors via API Import Injection}

\author{Juozas Dautartas\IEEEauthorrefmark{1}\orcidlink{0009-0000-9066-9565},~\IEEEmembership{Student Member,~IEEE,} Olga Kurasova\orcidlink{0000-0002-0570-1741},~\IEEEmembership{Senior Member,~IEEE,}
Juozapas Rokas Čypas\orcidlink{0009-0005-3844-3602}, ~\IEEEmembership{Student Member,~IEEE,}
and~Viktor Medvedev\orcidlink{0000-0001-5813-8545},~\IEEEmembership{Member,~IEEE}
\thanks{J.~Dautartas, J.~R.~Čypas, O.~Kurasova, and V.~Medvedev are with 
the Institute of Data Science and Digital Technologies, 
Faculty of Mathematics and Informatics, Vilnius 
University, LT-08412 Vilnius, Lithuania (e-mail: juozas.dautartas@mif.stud.vu.lt; \{juozapas.cypas, olga.kurasova, 
viktor.medvedev\}@mif.vu.lt).}%
\thanks{\IEEEauthorrefmark{1}Corresponding author: Juozas Dautartas.}%
\thanks{This work has been submitted to the IEEE for possible publication. Copyright may be transferred without notice, after which this version may no longer be accessible.}}


\markboth{Preprint}%
{Dautartas \MakeLowercase{\textit{et al.}}: Targeted Evasion of Malware Detectors via API Import Injection}

\IEEEpubid{}

\maketitle

\begin{abstract}
Machine learning-based malware detectors are widely deployed in antivirus and endpoint detection systems, yet their reliance on static features makes them vulnerable to adversarial manipulation.  This paper investigates whether a malware sample can be intentionally misclassified as a specific benign software category, not merely as ``not malware'', by adding a small number of Win32 API imports characteristic of that selected category, without removing any existing imports or retraining the detector. We propose a framework centered on a Conditional Variational Autoencoder (CVAE) whose decoder is strictly additive. It can introduce new API calls but never remove existing ones, preserving malware functionality by design. For each malware sample, the framework automatically identifies which benign category it most closely resembles and uses that as the evasion target. A knowledge-distilled differentiable proxy enables gradient-based training against the non-differentiable ensemble detector. Experiments on a six-class dataset of binary Win32 API import vectors extracted from 3,799 Windows executables (five benign categories, one malware class) show that, against a detector achieving 87.5\% malware recall, adding just 20 API imports reduces recall to 30\%. At $k=20$, among samples that evaded detection, 99\% are classified as the intended target category. The CVAE outperforms both a frequency-based baseline and random selection at every tested injection size ($k = 5$ to $50$). Validation on real PE files submitted to VirusTotal confirms that the attack transfers to commercial static detection engines, with an average 54.5\% reduction in flagging engines. These findings expose a concrete vulnerability in API-based malware classifiers and demonstrate that targeted evasion into a chosen benign category is achievable with minimal, functionality-preserving modifications.
\end{abstract}


\section{Introduction}
\label{sec:introduction}
 
\IEEEPARstart{M}{achine} learning (ML) has become a core component of modern malware detection. Antivirus and endpoint detection and response (EDR) systems increasingly rely on ML classifiers rather than traditional signature-based detection methods, as threats evolve faster than signatures can be written~\cite{Beyond_the_sandbox2025, malware_packers}.

One of the most widely used input features for these classifiers is the list of Win32 API functions imported by a piece of software. These imports are stored in the import address table (IAT) of portable executable (PE) files and reflect what operating system services the program intends to use. A program that imports file encryption or network tunneling functions looks very different from a text editor, and classifiers exploit this difference effectively.

However, this reliance on API imports for detection creates a practical weakness. Anyone can add new entries to a PE file's IAT without touching the existing code or removing any current imports. If the classifier interprets these new entries as signs of benign behavior, the prediction may shift away from malware. Prior work has demonstrated variants of this vulnerability through generative adversarial network (GAN)-based feature modification~\cite{BlackBoxGAN}, reinforcement learning over PE structure~\cite{anderson2018learning}, and byte-level injection~\cite{kolosnjaji2018adversarial, demetrio2021functionality}. AI-based security systems are both defensive tools and potential attack targets, so studying these weaknesses is valuable for improving defenses~\cite{Schroer2025}.

Despite these advances, nearly all existing malware evasion methods share a common limitation. They operate in a binary (malware vs.\ benign) setting and treat any misclassification away from the 
malware label as success. In practice, however, benign software is not a single undifferentiated category. Office tools, development software, multimedia players, security utilities, and administration tools all have distinct API usage patterns. A stronger and more realistic attack would force the classifier to predict a specific benign category, not just ``not malware''. Such targeted evasion achieves better results than evasion based on any benign class features~\cite{dautartas2026ieee}.

\begin{figure}[!t]
    \centering\includegraphics[width=0.94\columnwidth]{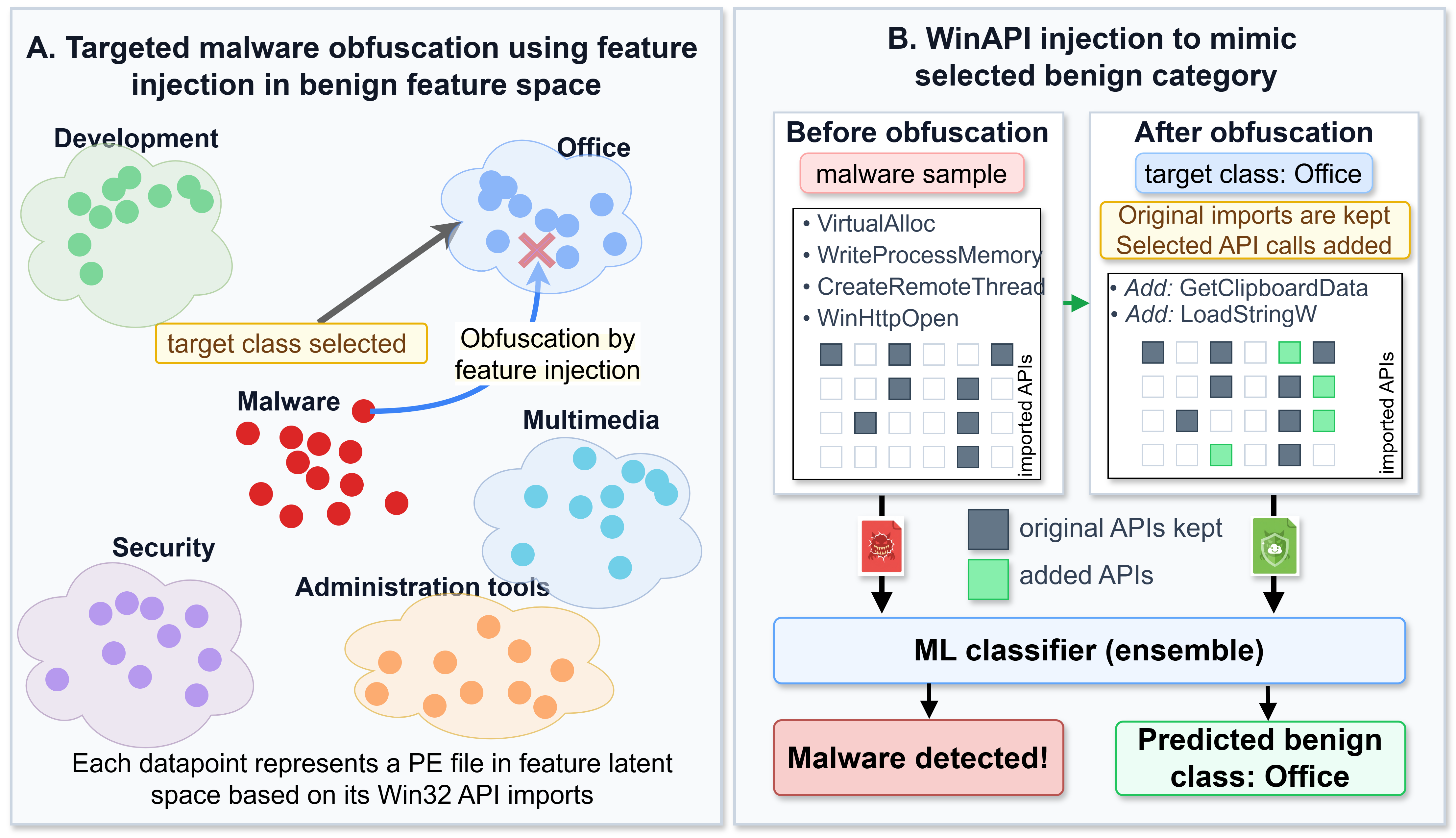}
    \caption{Conceptual overview of targeted malware evasion using API import injection.}
    \label{fig:concept}
\end{figure}

This paper investigates whether such targeted evasion is achievable by injecting a small number of category-specific Win32 API imports into malware samples, without removing any existing imports or retraining the detector.  Fig.~\ref{fig:concept} illustrates the core idea. In the feature space defined by API imports, software categories form distinct clusters (panel~A). By injecting a few API calls characteristic of a chosen benign category, a malware sample can be shifted from the malware cluster into the target benign region (panel~B).  No existing imports are removed, so malware functionality is preserved. The classifier, which correctly identified the sample before, now labels it as the chosen benign category. The central question addressed in this paper is whether such minimal, targeted additions can systematically force misclassification into a specific benign category, and whether the selection of which API calls to inject matters more than their quantity.

The main contributions of this paper are as follows:
\begin{enumerate}
  \item We introduce the concept of targeted adversarial evasion for API-based malware classifiers, where malware is not simply misclassified as benign, but as a specific chosen benign software category. To our knowledge, prior work in this domain has focused exclusively on untargeted evasion, treating all benign classes as a single undifferentiated target.
 
  \item We propose an attack framework based on a Conditional Variational Autoencoder (CVAE) with a strictly additive decoder. The decoder can only add API imports and never remove existing ones, preserving malware functionality by design. The framework also includes a strong ensemble-based malware detector as the evasion target and a knowledge-distilled differentiable proxy that enables gradient-based CVAE training.
  
  \item We construct and release a six-class dataset of binary Win32 API import vectors extracted from Windows executables, comprising five distinct benign software categories and one malware class. To our knowledge, this is the first publicly released dataset that supports category-targeted evasion research on Windows PE files.
\end{enumerate}

The remainder of this paper is organised as follows. Section~\ref{sec:relatedworks} reviews related work on malware detection and adversarial evasion. Section~\ref{sec:framework} describes the proposed framework. Section~\ref{sec:experiments} presents the experimental setup and results.  Section~\ref{sec:future} discusses  limitations and future directions, and Section~\ref{sec:conclusions} concludes the paper.

\section{Related Work} 
\label{sec:relatedworks}

This section reviews ML-based malware detection and the role of API features (Section~\ref{sec:rw_detection}), adversarial attacks on PE detectors (Section~\ref{sec:rw_attacks}), and generative evasion methods (Section~\ref{sec:rw_generative}). We conclude by identifying the research gap addressed in this work (Section~\ref{sec:rw_gap}).

\subsection{ML-Based Malware Detection}
\label{sec:rw_detection}

ML-based malware detection has been thoroughly studied using both static and dynamic features. Recent studies provide a comprehensive overview of AI-based detection methods and feature representations, including PE header and section statistics, string analysis, and behavioural traces extracted from dynamic analysis \cite{Yan2023}. In these studies, benign software is typically grouped into a single class, while malware is divided into several families or categories, such as viruses, worms, Trojan horses, ransomware, etc.

Commonly used datasets contain features extracted through static analysis, dynamic analysis, or a combination of both. Static analysis examines file properties such as entropy, section statistics, and IAT contents without executing the file~\cite{Malware_detect_rev}. Dynamic analysis executes the sample in an isolated sandbox to capture runtime behaviour such as system call sequences, network connections, and process relationships ~\cite{Malbert_malware_detection}. Among static features, Win32 API imports are particularly informative because they reveal what operating system services a program intends to use. API-MalDetect~\cite{Manihiro2023} show that API usage patterns alone are highly effective for malware detection. Explainability studies confirm that API-based features consistently rank among the strongest predictors in trained models~\cite{detectors_explainability,packing_detection_shap_vals}.

In most of this work, benign software is treated as a single class, while malware is divided into families such as ransomware, trojans, or worms. This design choice reflects the detection goal, but it also means that the distinct API patterns of different benign software categories remain largely unexplored. As ML-based detection became a core part of antivirus and EDR products, researchers naturally began investigating how these models can be attacked~\cite{Kozk2025,kolosnjaji2018adversarial}. 

\subsection{Adversarial Attacks on PE Malware Detectors}
\label{sec:rw_attacks}

Adversarial attacks on Windows PE malware detectors differ in two important ways~\cite{AdvAtkWinPE}. First, they differ in what is modified. Some attacks manipulate raw bytes or PE structure directly, while others operate in the feature space used by the classifier. Second, they differ in what is preserved. A modified sample may be valid according to the PE format specification but fail to execute, or it may execute correctly but no longer perform its original malicious actions, or it may retain its full malicious functionality unchanged. The last case is the strongest and most realistic constraint~\cite{AdvAtkWinPE}.

Attacks also differ in their goal. Untargeted attacks succeed as long as the detector no longer labels the sample as malware, regardless of what it predicts instead. Targeted attacks are more demanding, as the attacker seeks misclassification into a specific class, such as a particular category of benign software~\cite{Kucuk2020}. In the malware domain, this distinction matters. An attacker who can make malware appear as a trusted security tool or office application gains more than just evasion. In principle, such a sample could be treated with greater trust by the system, potentially bypassing security policies that distinguish between software types, though this remains to be verified empirically. 

\subsection{Generative Evasion Methods}
\label{sec:rw_generative}

Generative models are increasingly used to produce adversarial malware, as they can learn perturbations that systematically bypass detectors. One example is MalGAN~\cite{BlackBoxGAN}, which targets black-box malware detectors by training a generator against a substitute detector that approximates the decision behavior of the unknown target model. The generator learns to modify malware feature vectors so that the substitute detector no longer classifies them as malicious. MalGAN applies an additive constraint to preserve malware functionality. This functionality-preserving principle is a recurring requirement across adversarial malware methods, removing existing features risks breaking the malware's ability to execute~\cite{AndroidHIV2020}. Later GAN-based methods extend this idea to other feature representations, such as n-gram features extracted from executable bytecode~\cite{ZHU2022485}, or combine GANs with large language models to produce evasive variants~\cite{GAN_cleanware_features}.

Other methods manipulate the PE file more directly. Insertion-based attacks such as GAMMA~\cite{demetrio2021functionality} injects benign content into unused PE regions while balancing evasion success against payload size. Similarly, padding-based attacks on MalConv~\cite{kolosnjaji2018adversarial} append carefully chosen bytes to the end of the file to shift the detector's decision without changing the original code. GAPGAN~\cite{yuan2020black} follows a similar direction using a GAN to generate the appended payload.

Beyond GANs, reinforcement learning methods learn sequences of PE modifications that preserve executability while evading detection~\cite{anderson2018learning, RL_advgen}. Prototype-guided methods move samples toward a benign centroid in the detector's embedding space~\cite{Qiao2022}. Evolutionary and perturbation-based approaches have also demonstrated high evasion rates while preserving functional behaviour~\cite{FUMvar}. Realistic target-based adversarial attacks against MalConv and gradient-boosted models trained on Windows PE bytes is evaluated in~\cite{imran2024realistic}. 

Despite their variety, all of these methods share a common limitation. They operate in a binary malware versus benign setting and treat any misclassification away from malware as success, regardless of which benign class the sample lands in. This is significant because benign software is not a single homogeneous category. Office tools, development software, and security utilities all have distinct API usage patterns, and a detector that models these differences may behave very differently from one that treats them as a single class. Furthermore, most of these methods manipulate raw bytes or structural PE properties, making it difficult to reason about which specific benign category a modified sample resembles. API imports are a more natural feature space for this purpose, as different software categories exhibit characteristically different import patterns. To our knowledge, no prior work studies whether evasion can be directed  toward a specific chosen benign category. Table~\ref{tab:related_work_comparison} summarises these methods and highlights this gap.

A direct experimental comparison with these methods is not feasible, as all of them operate in a binary setting on different feature spaces such as raw bytes, PE structure, or learned embeddings, whereas our framework requires a multi-class dataset with distinct benign categories.  Instead, we compare the CVAE against two baselines that operate under identical conditions: a frequency-based strategy (MostPopular) and random selection, both described in Section~\ref{sec:setup}.

\begin{table*}[!t]
\centering
\small
\setlength{\tabcolsep}{4pt}
\renewcommand{\arraystretch}{1.15}
\caption{Comparison of adversarial malware evasion approaches.
  \#Cls: number of evaluation classes;
  Add.: strictly additive (no feature removal);
  Targ.: targeted evasion into a chosen benign category.}
\label{tab:related_work_comparison}
\begin{tabular}{@{}l l l c c c p{5.5cm}@{}}
\toprule
\textbf{Method}
  & \textbf{Approach}
  & \textbf{Feature space}
  & \textbf{\#Cls}
  & \textbf{Add.}
  & \textbf{Targ.}
  & \textbf{Key notes} \\
\midrule
MalGAN~\cite{BlackBoxGAN}
  & GAN + substitute detector
  & Binary static features
  & 2
  & \ding{51}
  & \ding{55}
  & Additive in feature vector; no demonstrated mapping to real PE modifications \\
Gym-Malware~\cite{anderson2018learning}
  & RL agent over PE actions
  & PE headers / sections
  & 2
  & \ding{51}
  & \ding{55}
  & Modifies real PE binaries; actions designed to preserve functionality \\
Padding~\cite{kolosnjaji2018adversarial}
  & Gradient-based byte append
  & Raw bytes
  & 2
  & \ding{51}
  & \ding{55}
  & Only appends bytes; original executable code unchanged \\
GAPGAN~\cite{yuan2020black}
  & GAN-generated payload
  & Raw bytes
  & 2
  & \ding{51}
  & \ding{55}
  & Only appends generated bytes; original executable code unchanged \\
GAMMA~\cite{demetrio2021functionality}
  & Constrained optimisation
  & Injected PE sections
  & 2
  & \ding{51}
  & \ding{55}
  & Injects benign content into unused PE regions; balances evasion vs.\ payload size \\
Prototype~\cite{Qiao2022}
  & Move toward benign centroid
  & Learned embeddings
  & 2
  & \ding{55}
  & \ding{55}
  & May add or remove features; single undifferentiated benign class \\
\midrule
\textbf{Ours}
  & \textbf{CVAE + target-selecting ensemble}
  & \textbf{Binary API imports}
  & \textbf{6}
  & \ding{51}
  & \ding{51}
  & \textbf{Distilled proxy for gradients; ArcFace embeddings; per-sample target selection} \\
\bottomrule
\end{tabular}
\end{table*}

\subsection{Research Gap}
\label{sec:rw_gap}
The methods reviewed above share three limitations that our work addresses. First, they operate in a binary setting, treating benign software as a single undifferentiated class and ignoring the distinct API profiles of different software categories. Second, targeted evasion, forcing misclassification into a specific benign category rather than simply away from malware, has received little attention in the malware domain. The few works in this direction operate on raw bytes rather than semantic features~\cite{imran2024realistic}, and although adding target class features is known to be an effective misclassification strategy~\cite{suciu2019exploring}, it has not been explored in a multi-class targeted setting. Third, most generative methods manipulate raw bytes or structural PE properties, making it difficult to reason about category-specific benign behaviours. API imports are a natural feature space for targeted evasion because different software categories exhibit characteristically different import patterns.

Our work addresses these gaps with a CVAE-based framework that injects category-specific Win32 API imports under a strictly additive constraint, preserving all original imports and thus malware functionality. Evaluated on a six-class dataset (five benign categories, one malware class), this is, to our knowledge, the first systematic study of category-aware API injection for targeted malware evasion.
\section{Proposed Framework}
\label{sec:framework}
 
This section describes the proposed targeted evasion framework. It comprises four components trained sequentially: (i)~ ensemble~A -- a multi-classifier ensemble trained on all classes, serving as the malware detector to be evaded (Section~\ref{sec:ensemble});
(ii)~proxy~$P$ -- a knowledge-distilled proxy that provides differentiable gradient signals (Section~\ref{sec:proxy});
(iii)~ensemble~B -- the same architecture as ensemble~A but trained on benign classes only, selecting the target class for each malware sample (Section~\ref{sec:ensemble}); and (iv)~CVAE -- a conditional variational autoencoder with an additive decoder that generates targeted perturbations (feature modifications) (Section~\ref{sec:cvae}). 

\begin{figure}[!t]
    \centering
    \includegraphics[width=0.90\columnwidth]{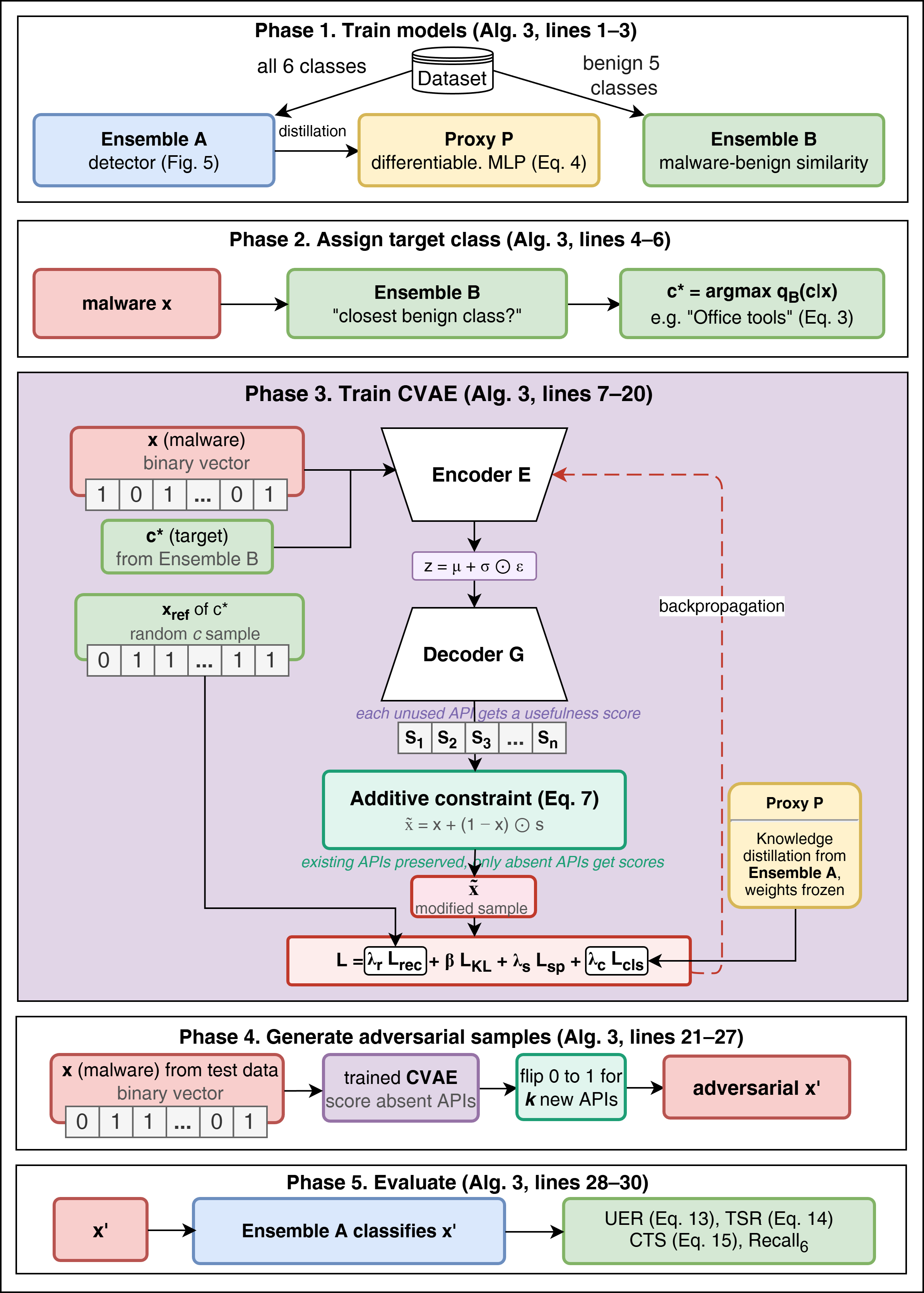}
    \caption{High-level overview of the proposed framework.}
    \label{fig:method_overview}
\end{figure}

Ensemble~B identifies which benign category each malware sample most closely resembles based on its API usage patterns. This target category then guides the CVAE to generate perturbations that shift the sample toward that specific benign region in the feature space. All perturbations are strictly additive: the framework can turn absent API calls to present ($0 \to 1$) but never removes existing ones, preserving the original malicious functionality by design. At evaluation time, only the top-$k$ highest-scored additions are applied, simulating a realistic attacker who can inject only a limited number of API calls. After perturbation, Ensemble~A is used to evaluate attack success by measuring how many malware samples are misclassified as the intended target category. The overall workflow is illustrated in Fig.~\ref{fig:method_overview} and formalised in Algorithm~\ref{alg:framework}. The framework operates under a grey-box threat model: the attacker knows the feature representation and has access to representative training data, but not the detector's internal parameters or weights, and is restricted to additive IAT modifications only. This constraint preserves malware functionality while limiting the attack surface to import injection only.

\subsection{Dataset and Feature Representation}
\label{sub:dataset}
A review of existing Windows malware datasets revealed that most malware datasets are unbalanced and have more malware samples compared to benign samples. Analyzed datasets have one class for benign software, while malware is categorized into more detailed groups based on malicious functionality such as ransomware, spyware, adware, etc.~\cite{dataset_fet_extr}. Other researchers split malware into specific malware families related to certain Advanced Persistent Threat (APT) activity. Observations reveal that these datasets are oriented towards malware detection and classification tasks. However, this can often lead to outdated datasets, as malware tends to evolve rapidly and implement new techniques, while others become outdated and detected. On the other hand, benign software tends to have more stable and clear features as it gets developed. Benign software authors have no need to evade detection or implement malicious functionality. Based on these observations, we have collected and presented a new dataset that is focused on benign software instead of malware. In the proposed dataset, malware is a single class collected from the MalwareBazaar \cite{malwarebazaar}, which is a widely used database of malware samples shared for research purposes. While benign software is grouped into five classes based on functionality and purpose of the software: productivity office tools (software and tools that are commonly used for everyday office tasks), development tools (IDEs and compilers), multimedia software (various software related to video or audio), security tools (various antivirus components and detection tools), and administration tools (software for regular system administration tasks) (see~Table~\ref{tab:dataset_summary}). The dataset used in this study is publicly available 
on Zenodo at DOI:~\href{https://doi.org/10.5281/zenodo.20208958}{10.5281/zenodo.20208958}~\cite{dautartas2026winapiadvmal}. For safety and legal reasons, the released dataset contains the extracted IAT-based feature representation required to reproduce the experiments, rather than redistributing raw malware binaries.

\begin{figure}[!t]
    \centering
    \includegraphics[width=0.91\columnwidth]{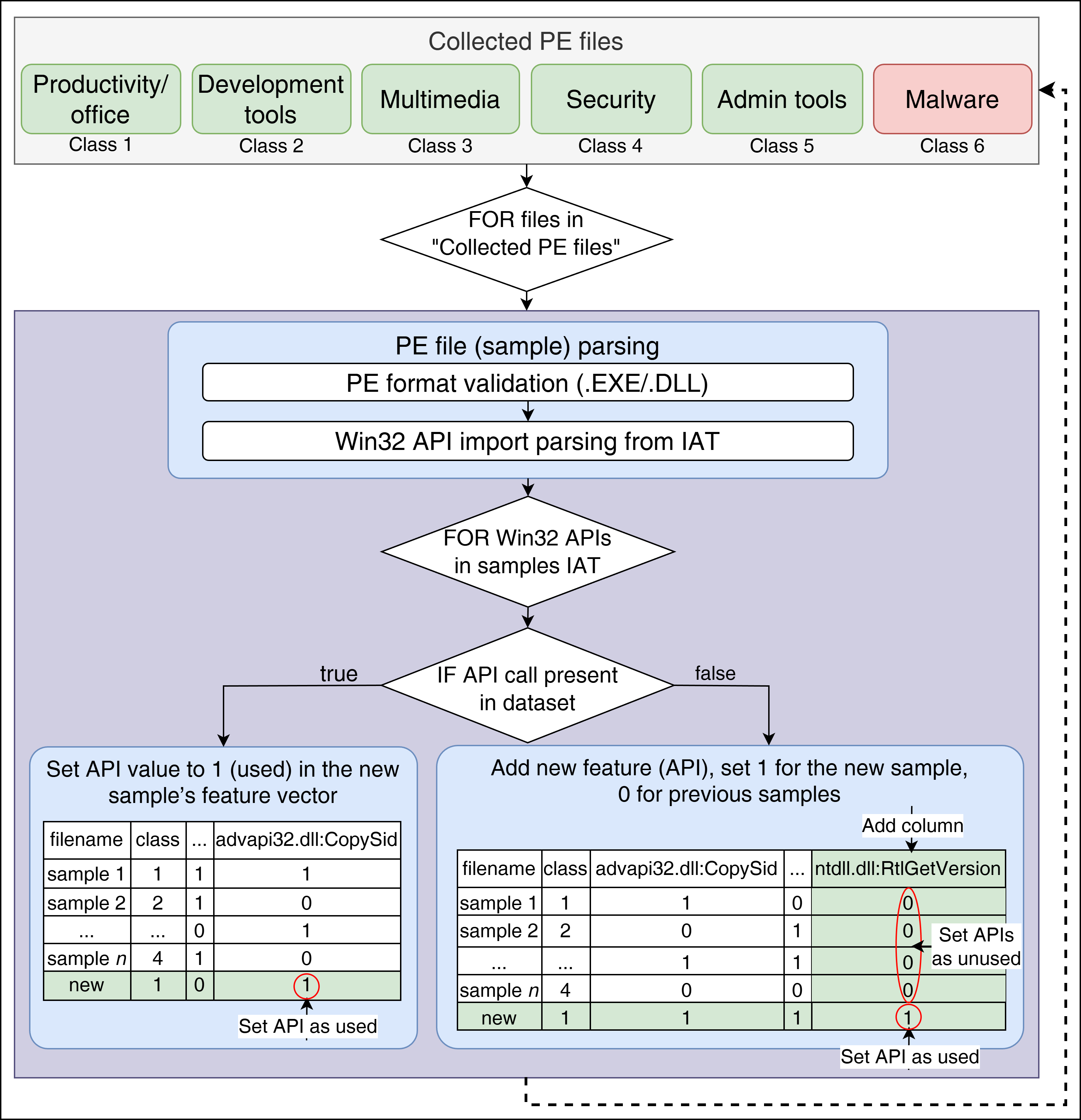}
    \caption{Dataset collection workflow.}
    \label{dataset_collection}
\end{figure}

\begin{table}
\centering
\caption{Dataset summary: class descriptions and per-class statistics
  of active API features.}
\label{tab:dataset_summary}
\begin{tabular}{cl r rrr}
\toprule
\textbf{Class} & \textbf{Category} & \textbf{Samples}
  & \textbf{Min} & \textbf{Max} & \textbf{Mean} \\
\midrule
1 & Office tools       & 667 &  5 & 1,268 &  69.0 \\
2 & Development tools  & 746 &  5 &   768  &  67.7 \\
3 & Multimedia         & 623 &  5 &   587  &  54.6 \\
4 & Security tools     & 519 &  6 &   792  & 125.0 \\
5 & Admin tools        & 566 &  5 &   877  & 148.5 \\
6 & Malware            & 678 &  5 &   673  & 111.4 \\
\midrule
  & \textbf{Total}     & \textbf{3,799} & & & \\
\bottomrule
\end{tabular}
\end{table}

The dataset collection pipeline is illustrated in Fig.~\ref{dataset_collection}. Benign samples were collected by installing standard Windows applications in each benign category and extracting PE files (.exe and .dll) directly from each installation. Malware samples were downloaded from MalwareBazaar database. Collected executable files were then parsed and analyzed. In this research, we focused on the IAT, which is part of static PE analysis. Extracted imports are added to dataset feature vector if not already present and marked as 1 (imported), other imports that are present in dataset and are not imported by specific sample are marked as 0 (not present in IAT for given file). Combining all observed values across the entire dataset defines a set of features, resulting in 2,713 binary features per sample. Samples with fewer than five imported functions were excluded.

\subsection{Ensemble-Based Malware Detector}
\label{sec:ensemble}

The framework requires a classifier that is both accurate enough to serve as a meaningful evasion target and robust enough to resist simple attack strategies. A single model may be vulnerable to 
specific weaknesses. For example, random forest (RF) captures non-linear feature interactions but may overfit to frequent patterns, while logistic regression (LR) generalises well on high-dimensional sparse data but assumes linear decision boundaries. Similarly, classifiers trained on raw binary features and on learned embeddings make different types of errors. By combining these complementary models, the ensemble reduces variance and fills in the gaps left by any single member. In the malware domain, such combinations can achieve higher detection rates and lower false positives than any standalone classifier~\cite{YOO2021}, and ensemble-based detectors have been shown to provide stronger resistance to adversarial attacks than single models~\cite{LiLi2020}. However, as our results 
demonstrate, they remain vulnerable to targeted feature injection.

\begin{figure}[tp]
    \centering
    \includegraphics[width=0.92\columnwidth]{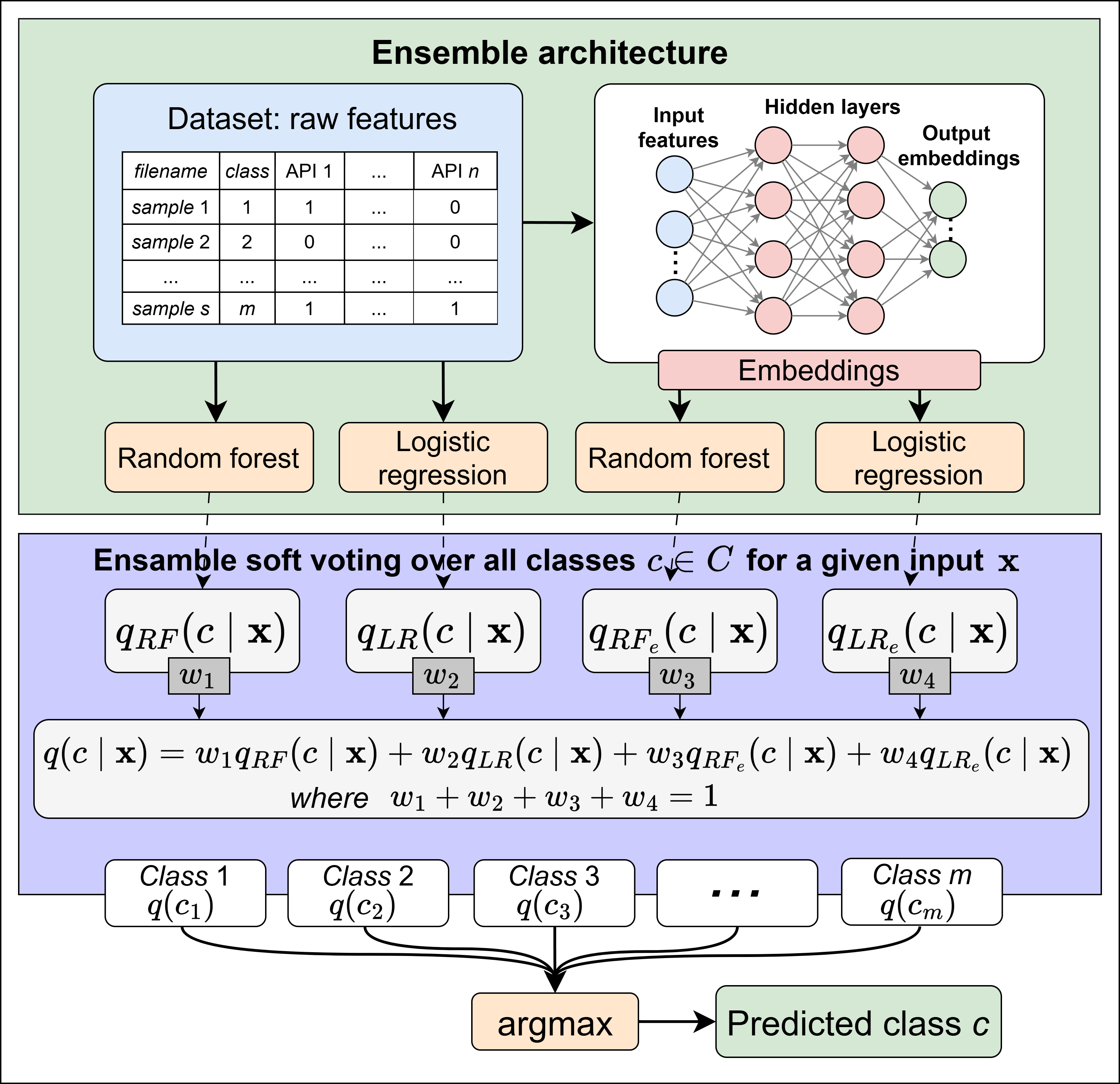}
    \caption{Ensemble training pipeline: RF and LR are trained on both raw features and learned embeddings; all classifiers are combined via weighted soft vote optimized on validation data.}
    \label{classifier_compare}
\end{figure}

The ensemble is built in four steps
(Algorithm~\ref{alg:ensemble} and Fig.~\ref{classifier_compare}):
 
\textit{Step 1}. RF and LR are trained on the preprocessed binary feature vectors (raw data). 

\begin{figure*}[!t]
    \centering
    \includegraphics[width=0.83\textwidth]{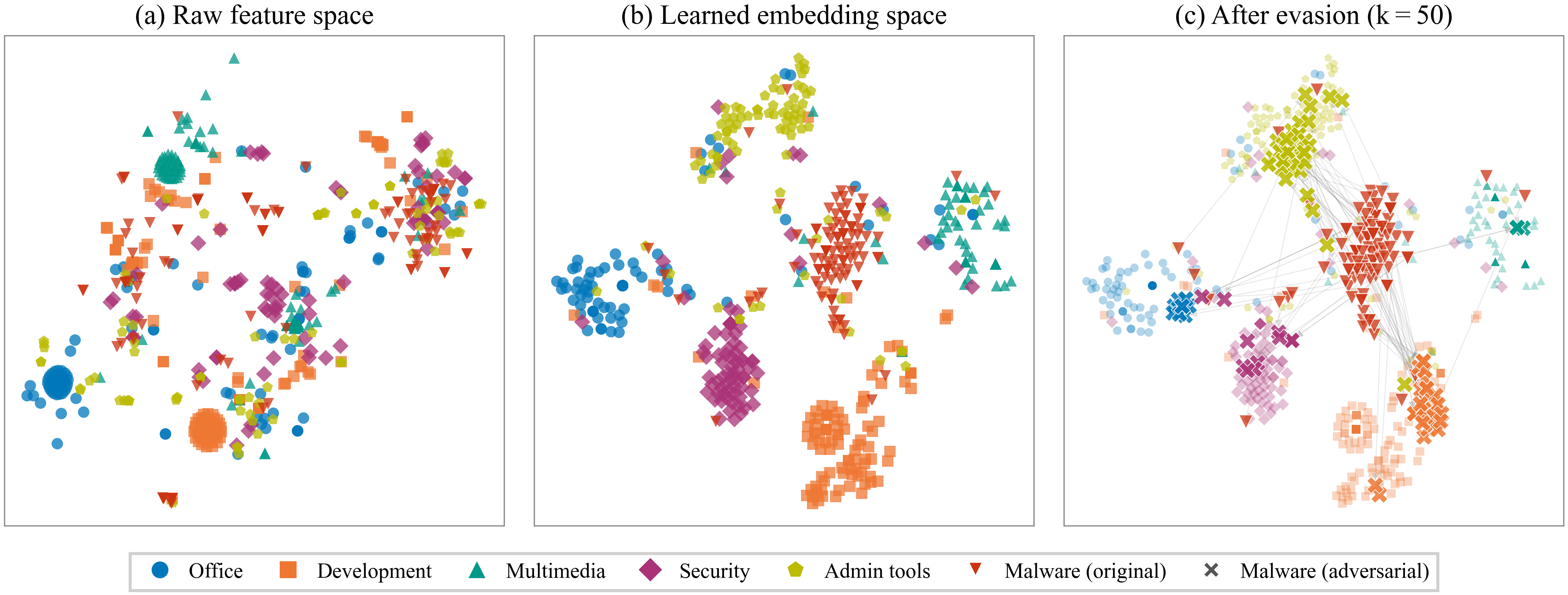}
    \caption{t-SNE projections of the test set at three stages of the framework: (a)~raw feature space, (b)~learned embedding space after encoder training, and (c)~embedding space after adversarial modification ($k{=}50$).}    \label{fig:tsne_panels}
\end{figure*}

\textit{Step 2}. As raw binary API vectors are high-dimensional (${\sim}$5,000 features) and sparse, classifiers trained directly on such vectors can overfit to superficial feature co-occurrences while missing deeper class structure. To obtain a more compact and discriminative representation, we train an MLP encoder with a residual block that maps preprocessed feature vectors into a 128-dimensional embedding space. The encoder architecture is provided in Table~\ref{tab:encoder_arch}. All embeddings are $L_2$-normalised. The encoder is trained with a combination of two loss functions. The first is cross-entropy (CE) with an ArcFace classification head~\cite{deng2019arcface}, which replaces standard logits with cosine similarities and adds an angular margin to the target class. This forces embeddings of the same class to cluster tightly and embeddings of different classes to separate by a guaranteed angular difference. This is in contrast to standard cross-entropy, which only requires that classes be linearly separable without constraining how compact or well-separated the clusters are. The second is the supervised contrastive (SupCon) loss~\cite{khosla2020supervised}, which directly pulls same-class embeddings closer and pushes different-class embeddings apart.  The combined loss function is:
\begin{equation}
\label{eq:encoder_loss}
L_{enc} = \mathrm{CE}(\text{ArcFace logits},\, \mathbf{y})
        + \lambda_{sc}\, L_{SupCon}(\mathbf{h},\, \mathbf{y}),
\end{equation}
where $\mathbf{h}$ denotes the $L_2$-normalised embeddings (hidden representation),
$\mathbf{y}$ are the class labels, and $\lambda_{sc} = 0.1$ weights the contrastive term. After training, the ArcFace classification head is discarded. The encoder is retained as a fixed feature extractor that maps any input to a 128-dimensional embedding. In the resulting embedding space, the separation of classes is much clearer than in the raw feature space, as visualised in Fig.~\ref{fig:tsne_panels}(a--b): the raw feature space shows considerable class overlap (panel~a), while the learned embeddings form compact, well-separated clusters (panel~b).

\begin{table}[!t]
  \centering
  \caption{MLP encoder architecture.}
  \label{tab:encoder_arch}
  \begin{tabular}{@{}lc@{}}
    \toprule
    \textbf{Layer} & \textbf{Output dim} \\
    \midrule
    FC + BN + ReLU + Dropout(0.3) & 1,024 \\
    FC + BN + ReLU + Dropout(0.3) & 512  \\
    \midrule
    \textit{Residual block:} & \\
    \quad FC + BN + ReLU + Dropout(0.3) & 512 \\
    \quad FC + BN                        & 512 \\
    \quad Add input + ReLU              & 512 \\
    \midrule
    FC (projection)              & 128  \\
    $L_2$ normalisation             & 128  \\
    \bottomrule
  \end{tabular}
\end{table}

\textit{Step 3}.
A second pair of RF and LR classifiers is trained on the
128-dimensional embeddings produced by the encoder from Step~2.
 
\textit{Step 4}.
Each of the four classifiers outputs a probability distribution over all classes $c \in \{1, \dots, C\}$, for a given input $\textbf{x}$ where $C$ is the number of classes. The ensemble combines them via a weighted soft vote:
\begin{equation}
\label{eq:ensemble}
\begin{split}
q(c \mid \textbf{x}) = w_1 q_{RF}(c\mid \textbf{x}) + w_2 q_{LR}(c\mid \textbf{x})\\
     + w_3 q_{RF_e}(c\mid \textbf{x}) + w_4 q_{LR_e}(c\mid \textbf{x}),
\end{split}
\end{equation}
where $q_m(c \mid \textbf{x})$ is the probability assigned to class $c$ by the $m$-th member, subscript~$e$ denotes embedding-based
classifiers, and the weights satisfy $\sum w_i = 1$.  The
weights are optimized on the validation set to maximize
macro-F1.  The predicted class is
$\hat{c} = \arg\max_{c}\, q(c\mid \textbf{x})$.

This architecture is used twice in the framework (see~Fig.~\ref{fig:method_overview}). Ensemble~A is trained on all six classes and serves as the malware detector to be evaded. Ensemble~B uses the same architecture but is trained on benign classes only. When presented with a malware sample, 
Ensemble~B identifies which benign category the sample most closely resembles in terms of API usage. That category is then used as the evasion target for the CVAE. The class with the highest probability is selected as the target:
\begin{equation}
\label{eq:target}
c^{*} = \arg\max_{c \in \{1,\dots,5\}} q_B(c \mid \textbf{x}),
\end{equation}
where $q_B(c)$ is the probability that ensemble~B assigns input \textbf{x} to class~$c$. This strategy is effective because the malware already partially resembles the selected target, so the CVAE only needs to bridge the remaining gap and requires fewer feature additions than if the target were chosen arbitrarily. Target classes are pre-computed once for all malware samples before CVAE training begins.

\subsection{Differentiable Proxy via Knowledge Distillation}
\label{sec:proxy}

The CVAE needs to learn which API calls to add by generating adversarial samples that fool ensemble~A, so gradients from the classification loss must propagate back through the decoder and encoder. However, the ensemble~A contains non-differentiable component such as RF that blocks gradient flow. Training the CVAE directly against the ensemble is therefore not possible. 

To address this, knowledge distillation~\cite{44873} was incorporated into the current methodology and used to train a differentiable proxy~$P$ that approximates the ensemble's predictions. In the distillation framework, the ensemble acts as the teacher and the proxy as the student, a smaller MLP (Table~\ref{tab:student-arch}) that learns to reproduce the output distribution of the ensemble. We refer to this model as a proxy rather than a student because its purpose is not to replace the ensemble as a standalone classifier, but solely to provide gradient signals on behalf of the non-differentiable ensemble during CVAE training.

\begin{table}[!t]
  \centering
  \caption{Proxy MLP architecture.}
  \label{tab:student-arch}
  \begin{tabular}{@{}lc@{}}
    \toprule
    \textbf{Layer} & \textbf{Output dim} \\
    \midrule
    FC + BN + ReLU + Dropout(0.1) & 2,048 \\
    FC + BN + ReLU + Dropout(0.1) & 1,024 \\
    FC + BN + ReLU + Dropout(0.1) & 512  \\
    FC (linear)                   & $C$  \\
    \bottomrule
  \end{tabular}
  \vspace{2pt}
  \\{\footnotesize FC: fully connected layer; BN: batch normalisation; $C$: number of classes.}
\end{table} 

The proxy is trained with a composite loss that combines hard labels (the ground-truth class) and soft labels (the ensemble's full probability distribution over all classes, which captures inter-class similarities that one-hot encodings discard): for example, if the ensemble assigns 0.6 probability to malware and 0.3 to security tools, the proxy learns that these two classes are relatively similar.
Before training begins, the teacher probability matrix is pre-computed by evaluating ensemble~A on the entire training set. The student is then trained for 700 epochs using Adam (learning rate $10^{-3}$) with the following loss:
\begin{equation}
\label{eq:distill_loss}
L_{\mathrm{distill}}
  = \alpha \cdot T^{2}\,\mathrm{KL}\!\bigl(
      \mathbf{q}_{\mathrm{soft}} \;\|\; \mathbf{p}_{\mathrm{soft}}
    \bigr)
  + (1 - \alpha) \cdot \mathrm{CE}\!\bigl(
      P(\mathbf{x}),\, y
    \bigr),
\end{equation}
where $\mathrm{KL}(\cdot \| \cdot)$ denotes the Kullback–Leibler divergence, $P(\mathbf{x})$ is the proxy's logit output, $y$ is the true class label, $\mathbf{p}_{soft}$ and $\mathbf{q}_{soft}$ are the softened distributions of the proxy and the ensemble respectively (defined below), $T$ is a temperature parameter, and $\alpha = 0.9$ so that the proxy primarily learns from the ensemble's distribution. The first term (KL divergence) teaches the proxy to reproduce how the ensemble distributes probability across all classes (not just the top prediction) capturing inter-class similarities that one-hot labels discard. The second term (cross-entropy) ensures the proxy also learns the correct class directly.


A temperature parameter $T{=}5$ is used to soften both distributions, spreading probability mass across classes to reveal inter-class similarities hidden in confident predictions. 

Because the proxy and the ensemble produce different types of output, each requires a different softening procedure. The proxy produces logits (unnormalised scores), which are softened by dividing by $T$ before applying softmax:
$p_{soft}(c\mid \textbf{x}) = \mathrm{Softmax}(P(\mathbf{x}) / T)_c$,
following~\cite{44873}, where $P(\mathbf{x})$ is the proxy's logit vector and subscript~$c$ selects the entry for class~$c$. The ensemble produces probabilities $q(c \mid \mathbf{x})$ directly (not logits),
so each probability is raised to the power $1/T$ and
renormalised:
$q_{soft}(c \mid \textbf{x}) = q(c \mid \textbf{x})^{1/T} \big/ \sum_{c'} q(c' \mid \textbf{x})^{1/T}$.
The $T^2$ factor in Eq.~\eqref{eq:distill_loss} compensates for the reduced gradient magnitude at higher temperatures.

After training, all weights of $P$ are frozen.  The proxy then serves as a differentiable stand-in for ensemble~A during CVAE training. Gradients from the classification loss flow through $P$ back into the CVAE encoder and decoder.

\subsection{Conditional Variational Autoencoder}
\label{sec:cvae}

Variational autoencoders (VAEs) encode data into a latent probability distribution from which new samples can be generated. However, standard VAEs lack a mechanism for specifying which class to generate, making them unsuitable for targeted evasion. Conditional VAEs (CVAEs)~\cite{sohn2015learning} address this by conditioning both encoding and decoding on a target class label. Combined with an additive output constraint, the CVAE can generate adversarial samples that impersonate a chosen benign class while preserving all original malware functionality.

The CVAE is the core generative component of the framework, assisted by the guidance ensemble~B for target selection and MLP model (distilled from ensemble A) for gradient-based training. Generated adversarial samples are evaluated against ensemble~A in a grey-box
scenario (see~Algorithm~\ref{alg:framework}).



\subsubsection{Architecture}
\label{sec:cvae_arch}

The CVAE takes a malware sample and a target benign class as input, and outputs a modified sample in which selected API calls have been added to make it resemble the target class. The process has two stages: encoding and decoding. The encoder $E$ receives a binary feature vector $\mathbf{x} \in \{0,1\}^n$ (the malware sample) and a target class label $c^{*}$ as input, where $n$ is the number of binary API features. The class label is first converted into a learnable embedding vector $\mathbf{e}_{c^{*}} \in \mathbb{R}^{d_e}$  ($d_e$ is the class embedding dimension), which is concatenated with the binary feature vector $\mathbf{x}$. This combined input passes through a three-layer MLP with LeakyReLU activations to produce two vectors: a mean $\boldsymbol{\mu} \in \mathbb{R}^d$, $d$ is the latent space dimension, and a log-variance $\log\boldsymbol{\sigma}^2 \in \mathbb{R}^d$. Together, they parameterise a Gaussian distribution in the latent space that encodes which modifications are needed to make~$\mathbf{x}$ resemble the target class~$c^{*}$. During training, a latent code is sampled via the reparameterisation trick \cite{kingma2014auto}:
\begin{equation}
\label{eq:reparam}
\mathbf{z} = \boldsymbol{\mu} + \boldsymbol{\sigma} \odot
    \boldsymbol{\epsilon}, \quad
    \boldsymbol{\epsilon} \sim \mathcal{N}(\mathbf{0}, \mathbf{I}).
\end{equation}

At inference time, the mean is used directly
($\mathbf{z} = \boldsymbol{\mu}$), yielding deterministic outputs. 

The decoder $G$ takes three inputs: the original source malware
sample $\mathbf{x} \in \{0,1\}^n$, the latent code
$\mathbf{z} \in \mathbb{R}^d$, and the target embedding vector $\mathbf{e}_{c^{*}} \in \mathbb{R}^{d_e}$ of the target class $c^{*}$. It processes them through a four-layer MLP and applies an element-wise sigmoid $\sigma(\cdot)$ to produce a score $s_j \in [0,1]$ for each of the $n$ features:
\begin{equation}
\label{eq:scores}
\mathbf{s} = \sigma\!\big(G(\mathbf{x},\, \mathbf{z},\,
    \mathbf{e}_{c^{*}})\big).
\end{equation}
Each score ($s_j$) indicates how useful it would be to add the $j$-th API call for impersonating the target class: values close to~1 are strong candidates, values close to~0 are unlikely to help. These scores are combined with the source via an additive constraint:
\begin{equation}
\label{eq:additive_decoder}
\tilde{x}_j = \begin{cases}
    1     & \text{if } x_j = 1 \quad \text{(preserved)}, \\
    s_j   & \text{if } x_j = 0 \quad \text{(candidate for addition)},
\end{cases}
\end{equation}
or equivalently, $\tilde{\mathbf{x}} = \mathbf{x} + (1 - \mathbf{x})
\odot \mathbf{s}$.  If $x_j = 1$ (the API call is already present), the factor $(1 - x_j) = 0$ blocks any change, the feature stays~1 regardless of the score.  If $x_j = 0$ (the API call is absent), the score
$s_j$ passes through, indicating whether this call should be added. This design guarantees that existing API calls are never removed, preserving malware functionality, while absent features receive a score indicating whether they should be added. 

In summary, existing API calls are always preserved (the
additive constraint forces their output to~1), while absent
features receive scores ranking their usefulness for
impersonating the target class.

\subsubsection{Training Objective}
\label{sec:cvae_loss}

The CVAE is trained with a composite loss:
\begin{equation}
\label{eq:cvae_loss}
L = \lambda_r L_{\mathrm{rec}} + \beta\, L_{\mathrm{KL}}
  + \lambda_s L_{\mathrm{sp}} + \lambda_c L_{\mathrm{cls}},
\end{equation}
where $\lambda_r, \beta, \lambda_s, \lambda_c$ are tunable weights. $L_{\mathrm{rec}}$ measures how closely the output matches real benign samples of the target class; $L_{\mathrm{KL}}$ keeps the latent space structured; $L_{\mathrm{sp}}$ limits the number of features added; and $L_{\mathrm{cls}}$ ensures the proxy classifies the output as the target class.

\paragraph{Reconstruction loss ($L_{\mathrm{rec}}$) -- ``Learn what benign software looks like''} During each training step, a reference sample $\mathbf{x}_{\mathrm{ref}}$
is drawn at random from the target benign class $c^{*}$, serving as a reconstruction target. The reconstruction loss measures how closely the decoder output
$\tilde{\mathbf{x}}$ matches~$\mathbf{x}_{\mathrm{ref}}$, using binary cross-entropy (BCE), computed only over features that are absent in the original malware sample ($x_j = 0$):
\begin{equation}
\label{eq:rec}
L_{\mathrm{rec}} = \frac{1}{n_0}
  \sum_{\{j\,:\, x_j = 0\}}
  \mathrm{BCE}\!\big(\tilde{x}_j,\; x_{\mathrm{ref},j}\big),
\end{equation}
where $n_0 = |\{j : x_j = 0\}|$. Features already present in~$\mathbf{x}$ ($x_j = 1$) are excluded because the additive constraint fixes their output to~1, making their gradients uninformative.  By masking the loss to absent features only, the model learns which API calls are characteristic of each benign class and should be added.

\paragraph{KL divergence ($L_{\mathrm{KL}}$) -- ``Keep the latent space structured''}
The KL divergence term prevents the encoder from memorising individual training samples by encouraging the learned posterior $q(\mathbf{z}\mid\mathbf{x})$ to stay close to a standard Gaussian prior $\mathcal{N}(\mathbf{0}, \mathbf{I})$. Without this constraint, the encoder could assign each sample an isolated point in the $d$-dimensional latent space, making it unable to produce meaningful outputs for new, unseen malware samples:
\begin{equation}
\label{eq:kl}
L_{\mathrm{KL}} = -\frac{1}{2}\sum_{i=1}^{d}
  \bigl(1 + \log\sigma_i^{2} - \mu_i^{2} - \sigma_i^{2}\bigr),
\end{equation}
where $\boldsymbol{\mu}$ and $\boldsymbol{\sigma}^2$ are the mean and variance vectors produced by the encoder, and $d$ is the latent dimension. This prevents overfitting and improves generalisation to unseen malware samples. 

\paragraph{Sparsity penalty ($L_{\mathrm{sp}}$) -- ``Be selective about which features to add''} The decoder outputs scores for all $n$ features, but at evaluation time only $k$ of them will be selected and added.  If the decoder assigns similar scores to many features, it becomes unclear which ones are truly important for impersonating the target class. The sparsity penalty addresses this by penalising the total number of features the decoder proposes to add:
\begin{equation}
\label{eq:sparsity}
L_{\mathrm{sp}} = \sum_{j=1}^{n}
  \bigl(\tilde{x}_j - x_j\bigr),
\end{equation}
where $n$ is the number of features, $\tilde{x}_j$ is the decoder output, $x_j$ is the original value. Since the additive constraint guarantees $\tilde{x}_j \geq x_j$, all differences are non-negative and no absolute value is needed.  This penalty pushes the model to give high scores to a small number of features and near-zero scores to the rest, so that the $k$ selected features are chosen with higher confidence.

\paragraph{Classification loss ($L_{\mathrm{cls}}$) -- ``Actually
fool the classifier''} The three losses above shape the decoder's output to be sparse, structured, and similar to real benign samples, but none of them directly verifies whether the result actually fools a classifier. The classification loss addresses this by passing the modified sample through the frozen proxy~$P$ and measuring how confidently it assigns the target class~$c^{*}$.

Since $P$ was trained on binary $\{0,1\}$ inputs, the continuous decoder output must be rounded before being passed through it. However, rounding is not differentiable, which
would block gradient flow.  We use a straight-through estimator~\cite{bengio2013estimating}: the forward pass rounds to $\{0,1\}$, while the backward pass propagates gradients through the continuous scores as if rounding had not occurred.  The loss is:
\begin{equation}
\label{eq:cls_loss}
L_{\mathrm{cls}} = \mathrm{CE}\!\big(P(\tilde{\mathbf{x}}_{\mathrm{bin}}),
    \; c^{*}\big),
\end{equation}
where $\tilde{\mathbf{x}}_{\mathrm{bin}}$ is the binarised decoder output and $\mathrm{CE}$ denotes cross-entropy.

\paragraph{Hyperparameter tuning and training}

All loss weights $(\lambda_r, \beta, \lambda_s, \lambda_c)$, where $\beta$ follows the $\beta$-VAE
approach~\cite{higgins2017beta}, architectural
hyperparameters (latent dimension~$d$, embedding
dimension~$d_e$), and learning rate are jointly tuned using Optuna~\cite{akiba2019optuna} with a TPE sampler.  The tuning objective is a weighted combination of evasion metrics (Section~\ref{sec:metrics}) evaluated on class-6 samples from~$D_{val}$, directly measuring how effectively the CVAE fools the detector.

After hyperparameters are selected, the CVAE is retrained
from scratch using a separate held-out set~$D_{val\_es}$ for early stopping, so that model selection does not reuse data already seen during hyperparameter tuning.  The encoder and decoder are optimised jointly with Adam ($\beta_1{=}0.5$, $\beta_2{=}0.999$) and gradient clipping at norm~5. The complete CVAE training procedure is given in  Algorithm~\ref{alg:framework}.

\subsection{Top-$k$ Additive Feature Injection}
\label{sec:topk}

\begin{figure}[!t]
    \centering
    \includegraphics[width=0.90\columnwidth]{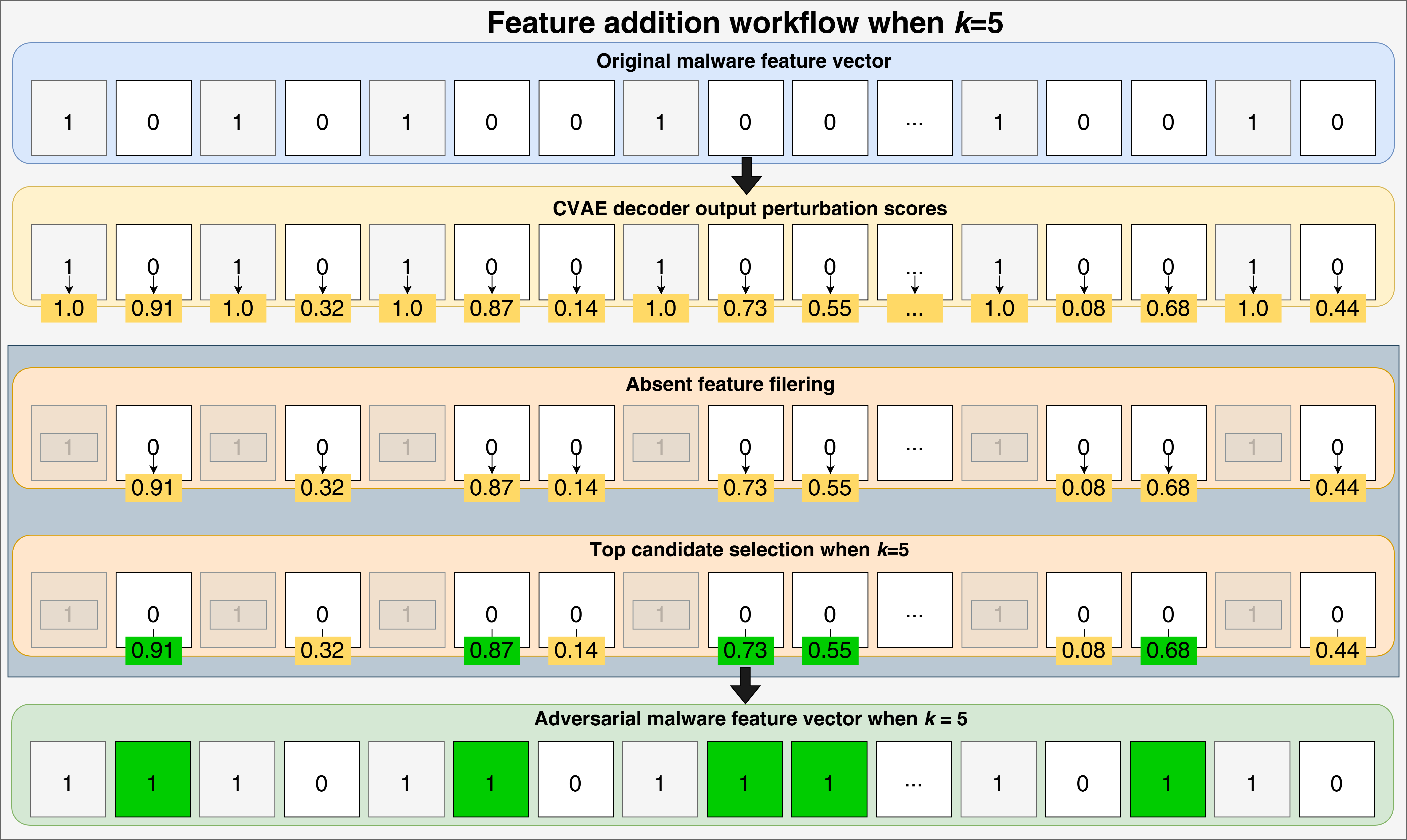}
    \caption{Feature addition workflow for $k{=}5$. The decoder scores all features; only absent features ($x_j=0$) are candidates; the five highest-scored are flipped to~1 (shown in green).}
    \label{number_of_k}
\end{figure}

In practice, an attacker can only inject a limited number of additional API imports into a PE file. Adding all high-scoring features at once would introduce too many changes and make the modification easy to detect. To simulate this constraint, we allow exactly $k$ new API calls to be added per sample. All features already present in the original sample ($x_j = 1$) are preserved unchanged.  Among the absent features ($x_j = 0$), we rank them by their decoder scores $s_j$ (Eq.~\ref{eq:scores}) and select the top~$k$.  These $k$ features are flipped from~0 to~1, producing the final adversarial sample.  Fig.~\ref{number_of_k} illustrates this process for $k{=}5$.

\section{Experiments and Analysis}
\label{sec:experiments}

\subsection{Experimental Setup}
\label{sec:setup}
Experiments use the dataset described in
Section~\ref{sub:dataset}. The data is split into four non-overlapping subsets: training (70\%), validation (10\%) for Optuna hyperparameter tuning, validation (5\%) for CVAE early stopping, and test (15\%) for final evaluation. The splits are stratified to preserve class proportions.

The entire pipeline (Fig.~\ref{fig:method_overview}) is repeated independently 20~times with different random initialisations. All metrics (Section~\ref{sec:metrics}) are computed on the held-out test set of each run and reported as mean and standard deviation.  All methods are evaluated across $k \in \{5, 10, 15, \dots, 50\}$ added API calls.

Existing adversarial malware evasion methods
(Table~\ref{tab:related_work_comparison}) operate in a binary malware-vs-benign setting and are not designed for targeted evasion into a specific benign category.  A direct comparison
would therefore not be meaningful, as these methods optimise for a different objective.  Instead, we compare the proposed CVAE-based framework against two baselines that use the same target class assignment (from ensemble~B):

\begin{itemize}
  \item MostPopular: for each malware sample, the $k$~API calls that occur most frequently among training samples of the assigned target class~$c^{*}$ are selected. This represents a simple, non-learned strategy that exploits class-level statistics \cite{dautartas2026ieee}.
  \item Random: $k$~absent API calls are selected uniformly at random, regardless of the target class. This serves as a lower bound, measuring whether arbitrary   additions have any evasion effect.
\end{itemize}

All experiments were conducted on a single NVIDIA GeForce RTX 4070 GPU with 8 GB of memory, 64 GB RAM. The framework was implemented in PyTorch.
 
\begin{algorithm}[t]
\caption{Ensemble classifier training}
\label{alg:ensemble}
\begin{algorithmic}[1]
\Require Train $D_{train}$ and validation $D_{val}$ data, class~set~$C$
\Ensure Ensemble classifier $F_{ens}$

\Statex \textbf{$\triangleright$ Step 1. Train classifiers on raw features}
\State Train RandomForest $RF$ and LogisticRegression $LR$
       on $D_{train}$

\Statex \textbf{$\triangleright$ Step 2. Train embedding encoder}
\State Train $MLPEncoder$ on $D_{train}$
       \Comment{Eq.~\eqref{eq:encoder_loss}}
\State $X_{e} \gets MLPEncoder(D_{train})$

\Statex \textbf{$\triangleright$ Step 3. Train classifiers on embeddings}
\State Train $RF_e$ and $LR_e$ on $X_{e}$

\Statex \textbf{$\triangleright$ Step 4. Optimise ensemble weights}
\State Find weights $(w_1, w_2, w_3, w_4)$ with $\sum w_i = 1$ that maximise macro-F1 on $D_{val}$
\Statex \Comment{$w_i$ is a contribution of the $i$-th ensemble member}
\State Define the ensemble probability for class $c \in C$
\Statex $q(c \mid \mathbf{x}) \gets w_1\, q_{RF}(c \mid \mathbf{x}) + w_2\, q_{LR}(c \mid \mathbf{x})+ w_3\, q_{RF_e}(c \mid \mathbf{x}) + w_4\, q_{LR_e}(c \mid \mathbf{x})$
\State \Return $F_{ens}$ \Comment{use as $\hat{c} = \arg\max_{c \in C}\, q(c \mid \mathbf{x})$}
\end{algorithmic}
\end{algorithm}

\begin{algorithm}[t]
   \caption{Proxy training via knowledge distillation}
    \label{alg:distill}
    \begin{algorithmic}[1]
    \Require Train data $D_{train}$ with labels $Y_{train}$, class set $C$, trained ensemble $F_{ens}$ (teacher), temperature $T$, soft-label weight $\alpha$
    \Ensure Differentiable proxy $P$ (student)
    
\Statex \textbf{$\triangleright$ Step 1. Collect soft targets from teacher}
    \State $Q \gets F_{\mathrm{ens}}(D_{train})$ 
    \Comment{Each row  $\mathbf{q}$ of $Q$ is a probability vector over classes from $C$}
\Statex \textbf{$\triangleright$ Step 2. Train proxy network}
    \State Initialize MLP network $P$
    \For{each training epoch}
     \For{each sample $(\mathbf{x}, y, \mathbf{q})$
    from $(D_{train}, Y_{train}, Q)$}
        \State $L_{hard} \gets \mathrm{CE} (P(\mathbf{x}), \; y)$
        \State $\mathbf{q}_{soft} \gets \mathbf{q}^{1/T} \big/ {\textstyle\sum_{c'} q_{c'}^{1/T}}$
      \Comment{soften teacher, element-wise}
         \State $\mathbf{p}_{soft} \gets \text{Softmax}(P(\mathbf{x})\;/\;T)$ \Comment{soften proxy}
        \State $L_{soft} \gets \text{KL}(\mathbf{q}_{soft} \;\|\; \mathbf{p}_{soft}) \cdot T^2$
        \State $L_{\mathrm{distill}} \gets \alpha \cdot L_{soft} + (1 - \alpha) \cdot L_{hard}$
        \Comment{Eq.~\ref{eq:distill_loss}}
        \State Update $P$ via Adam  on $L_{distill}$
     \EndFor
    \EndFor
    
\Statex \textbf{$\triangleright$ Step 3. Freeze and return}
    \State Freeze all weights of $P$
    \State \Return $P$
    \end{algorithmic}
\end{algorithm}

\begin{algorithm}[t]
\caption{Targeted malware evasion framework}
\label{alg:framework}
\begin{algorithmic}[1]
\Require Train $D_{train}$, validation $D_{val}$, $D_{val\_es}$, and $D_{test}$ data
\Comment{with benign classes $1,\dots,5$ and malware class $6$}
\Ensure Adversarial samples $X_{adv}$; evasion metrics
\Statex \textbf{$\triangleright$ Phase 1. Train models}
\State Train \textbf{Ensemble~A} on $D_{train}$ using all 6 classes \Comment{Alg.~\ref{alg:ensemble}}
\State Train \textbf{Ensemble~B} on $D_{train}$ using benign classes 1--5 only \Comment{Alg.~\ref{alg:ensemble}}
\State Train \textbf{Proxy}~$P$ by distilling \textbf{Ensemble~A}; freeze $P$ \Comment{Alg.~\ref{alg:distill}}
\Statex \textbf{$\triangleright$ Phase 2. Pre-compute target class for all malware samples}
\For{each malware sample $\mathbf{x}$ in $D_{train}$, $D_{val}$, $D_{val\_es}$, $D_{test}$}
\State $c^{*} \gets \arg\max_{c \in \{1,\dots,5\}} q_B(c \mid \mathbf{x})$
           \Comment{Eq.~\ref{eq:target}}
\EndFor
\Statex \textbf{$\triangleright$ Phase 3. Tune and train CVAE}
\State Tune hyperparameters via Optuna on class-6 samples from $D_{val}$
\State Train \textbf{Encoder} $E$ and \textbf{Decoder} $G$ on $D_{train}$ with early stopping on $D_{val\_es}$ 
\For{epoch $= 1, \dots, N_{epochs}$}
    \For{each malware sample $\mathbf{x}$ in $D_{train}$ with target $c^{*}$}
        \State Sample $\mathbf{x}_{ref}$ from benign class $c^{*}$ \Comment{for $L_{rec}$}
        \State $(\boldsymbol{\mu}, \log\boldsymbol{\sigma}^2) \gets E(\mathbf{x}, c^{*})$
        \State $\mathbf{z} \gets \boldsymbol{\mu} + \boldsymbol{\sigma} \odot \boldsymbol{\epsilon}, \;\; \boldsymbol{\epsilon} \sim \mathcal{N}(\mathbf{0}, \mathbf{I})$ \Comment{Eq.~\ref{eq:reparam}}
        \State $\mathbf{s} \gets \sigma\!\big(G(\mathbf{x},\, \mathbf{z},\, \mathbf{e}_{c^{*}})\big)$ \Comment{$\mathbf{e}_{c^{*}}$ is learnable class embedding, Eq.~\ref{eq:scores}}
        \State $\tilde{\mathbf{x}} \gets \mathbf{x} + (1 - \mathbf{x}) \odot \mathbf{s}$ \Comment{Eq.~\ref{eq:additive_decoder}}
        \State $L_{cls} \gets \mathrm{CE}\!\big(P(\tilde{\mathbf{x}}_{\mathrm{bin}}),\, c^{*}\big)$ \Comment{Eq.~\ref{eq:cls_loss}}
        \State $L \gets \lambda_r L_{rec} + \beta L_{KL} + \lambda_s L_{sp} + \lambda_c L_{cls}$ \Comment{Eq.~\ref{eq:cvae_loss}}
        \State Update $E, G$ on $L$
    \EndFor
\EndFor
\Statex \textbf{$\triangleright$ Phase 4. Generate adversarial samples}
\State $X_{adv} \gets \emptyset$
\For{each malware sample $\mathbf{x}$ in $D_{test}$ with target $c^{*}$}
    \State $(\boldsymbol{\mu}, \_) \gets E(\mathbf{x},\, c^{*})$ \Comment{use mean only at inference}
    \State $\mathbf{s} \gets \sigma\!\big(G(\mathbf{x},\, \boldsymbol{\mu},\, \mathbf{e}_{c^{*}})\big)$ \Comment{$\mathbf{z}=\boldsymbol{\mu}$, no sampling at inference}
    \State Select $k$ indices from $\{j \mid x_j = 0\}$ with highest $s_j$
    \State $\mathbf{x}' \gets \mathbf{x}$; set $x'_j \gets 1$ for selected; add $\mathbf{x}'$ to $X_{adv}$
\EndFor
\Statex \textbf{$\triangleright$ Phase 5. Evaluate on $D_{test}$}
\State Classify $X_{adv}$ with \textbf{Ensemble~A}
\State Compute UER (Eq.~\ref{eq:uer}), TSR (Eq.~\ref{eq:tsr}), CTS (Eq.~\ref{eq:cts})
\State \Return $X_{adv}$, UER, TSR, CTS
\end{algorithmic}
\end{algorithm}

\subsection{Evaluation Metrics}
\label{sec:metrics}

Evasion performance is measured using three metrics, all computed over the class-6 (malware) samples in the test set. Let $m_{malware}$ denote the total number of such samples. After applying adversarial perturbations, each sample is reclassified by ensemble~A, producing two counts: $m_{evaded}$ (number of malware samples no longer classified as malware and assigned to any benign class), $m_{target}$ (number of malware samples no longer classified as malware and classified as the intended target class $c^{*}$).

The experiments aim to answer three complementary questions about the proposed attack: (1)~Can CVAE cause the detector to miss malware at all? (2)~When this happens, does the sample fall into the specific benign class chosen by the reference ensemble~B? (3)~How many API imports are needed to achieve this reliably? Each question is addressed by a dedicated metric.

The untargeted evasion rate (UER) measures the fraction of malware samples that are no longer classified as malware, regardless of which benign class they receive:
\begin{equation}
\label{eq:uer}
\mathrm{UER} = \frac{m_{evaded}}{m_{malware}}\,.
\end{equation}
The targeted success rate (TSR) measures the fraction of malware samples classified as the intended target class $c^{*}$:
\begin{equation}
\label{eq:tsr}
\mathrm{TSR} = \frac{m_{target}}{m_{malware}}\,.
\end{equation}

The conditional target success (CTS) measures, among samples that successfully evaded the malware label, the fraction that landed on the intended target class rather than another benign class:
\begin{equation}
\label{eq:cts}
\mathrm{CTS} = \frac{m_{target}}{m_{evaded}}
             = \frac{\mathrm{TSR}}{\mathrm{UER}}\,.
\end{equation}

Since every sample counted in $m_{target}$ must also have evaded class-6, $\mathrm{TSR} \leq \mathrm{UER}$ always holds. High UER with low CTS indicates that evaded samples scatter across multiple benign classes, whereas high CTS indicates precise, targeted evasion. We additionally report class-6 recall after the attack from the defender's perspective: $\mathrm{Recall}_6 = 1 - \mathrm{UER}$. While UER, TSR, and CTS quantify the attacker's success, class-6 recall directly answers the practical question a defender faces, i.e. what proportion of malware would the detector still catch. We use class-6 recall as the primary metric in our analysis for this reason.

To verify that ensemble~A is a competent classifier worth attacking, we also report standard multi-class 
metrics: accuracy, macro-F1, macro-recall, and $\mathrm{Recall}_6$ before any attack.  

\subsection{Classification Performance Before Attack}
\label{sec:ensemble_results}

Before analysing evasion results, we verify that the malware detector and its differentiable proxy are competent classifiers. Table~\ref{tab:classifier_comparison} compares the ensemble with its individual members on the test set, reporting accuracy, macro-F1, macro-recall, and $\mathrm{Recall}_6$ (recall on the malware class). The ensemble achieves the best score on every metric, confirming that combining classifiers trained on raw and embedded features is genuinely complementary rather than redundant. The malware-class recall of 0.879 defines the baseline that the CVAE attack must overcome. The differentiable proxy trained via knowledge distillation (Section~\ref{sec:proxy}) attains $0.846 \pm 0.015$ test accuracy, only 0.7 percentage points below the ensemble it imitates, which is sufficient to provide informative gradient signals during CVAE training while remaining differentiable.

\begin{table}[!t]
\centering
\caption{Six-class classification performance, averaged over 20 runs. RF, LR trained on raw binary API features; RF + Emb, LR + Emb trained on the encoder's embeddings; Ensemble reflects a weighted soft vote of the four members (Eq.~\ref{eq:ensemble}).}
\label{tab:classifier_comparison}
\begin{tabular}{lcccc}
\toprule
\textbf{Model} & \textbf{Acc.} & \textbf{F1} & \textbf{Recall} & $\textbf{Recall}_6$ \\
\midrule
RF             & 0.845 & 0.844 & 0.845 & 0.869 \\
LR             & 0.829 & 0.826 & 0.825 & 0.871 \\
RF + Emb       & 0.844 & 0.842 & 0.842 & 0.871 \\
LR + Emb       & 0.844 & 0.842 & 0.841 & 0.879 \\
\textbf{Ensemble} & \textbf{0.853} & \textbf{0.851} & \textbf{0.851} & \textbf{0.879} \\
\bottomrule
\end{tabular}
\end{table}

\subsection{Evasion Results}
\label{sec:evasion_results}

Table~\ref{tab:comparison} summarises the evasion results for the CVAE, MostPopular, and Random methods across $k \in \{5, \dots, 50\}$ injected API calls. All values are averaged over 20~runs with different random initialisations.

\begin{table*}[!t]
\centering
\caption{Evasion performance vs.\ number of injected API calls $k$,
  averaged over 20 runs (mean\,$\pm$\,std).
  $\mathrm{Recall}_6$ reflects the defender's perspective: lower values indicate a more effective attack.
  At $k{=}0$, $\mathrm{Recall}_6 = 0.8745\pm0.05$ for all methods.
  The best result per $k$ is shown in bold.}
\label{tab:comparison}
\renewcommand{\arraystretch}{1.1}
\resizebox{\textwidth}{!}{%
\begin{tabular}{c cccc cccc cccc}
\toprule
& \multicolumn{4}{c}{\textbf{CVAE (Ours)}}
& \multicolumn{4}{c}{\textbf{MostPopular}}
& \multicolumn{4}{c}{\textbf{Random}} \\
\cmidrule(lr){2-5}\cmidrule(lr){6-9}\cmidrule(lr){10-13}
$k$ & UER & TSR & CTS & Recall$_6$
    & UER & TSR & CTS & Recall$_6$
    & UER & TSR & CTS & Recall$_6$ \\
\midrule
 5 & \textbf{0.2505\,$\pm$\,0.08} & \textbf{0.2407\,$\pm$\,0.09} & \textbf{0.9553\,$\pm$\,0.06} & \textbf{0.7495\,$\pm$\,0.08} & 0.1627\,$\pm$\,0.05 & 0.1387\,$\pm$\,0.05 & 0.8471\,$\pm$\,0.12 & 0.8373\,$\pm$\,0.05 & 0.1279\,$\pm$\,0.04 & 0.1108\,$\pm$\,0.04 & 0.8685\,$\pm$\,0.10 & 0.8721\,$\pm$\,0.04 \\
10 & \textbf{0.4118\,$\pm$\,0.18} & \textbf{0.3990\,$\pm$\,0.18} & \textbf{0.9625\,$\pm$\,0.05} & \textbf{0.5882\,$\pm$\,0.18} & 0.2260\,$\pm$\,0.06 & 0.1971\,$\pm$\,0.07 & 0.8650\,$\pm$\,0.09 & 0.7740\,$\pm$\,0.06 & 0.1289\,$\pm$\,0.04 & 0.1103\,$\pm$\,0.04 & 0.8586\,$\pm$\,0.11 & 0.8711\,$\pm$\,0.04 \\
15 & \textbf{0.5647\,$\pm$\,0.20} & \textbf{0.5559\,$\pm$\,0.20} & \textbf{0.9828\,$\pm$\,0.03} & \textbf{0.4353\,$\pm$\,0.20} & 0.2618\,$\pm$\,0.05 & 0.2284\,$\pm$\,0.05 & 0.8724\,$\pm$\,0.09 & 0.7382\,$\pm$\,0.05 & 0.1250\,$\pm$\,0.04 & 0.1034\,$\pm$\,0.04 & 0.8270\,$\pm$\,0.13 & 0.8750\,$\pm$\,0.04 \\
20 & \textbf{0.7005\,$\pm$\,0.18} & \textbf{0.6961\,$\pm$\,0.18} & \textbf{0.9933\,$\pm$\,0.01} & \textbf{0.2995\,$\pm$\,0.18} & 0.3103\,$\pm$\,0.05 & 0.2828\,$\pm$\,0.06 & 0.9089\,$\pm$\,0.06 & 0.6897\,$\pm$\,0.05 & 0.1279\,$\pm$\,0.05 & 0.1059\,$\pm$\,0.04 & 0.8403\,$\pm$\,0.12 & 0.8721\,$\pm$\,0.05 \\
30 & \textbf{0.8417\,$\pm$\,0.17} & \textbf{0.8343\,$\pm$\,0.16} & \textbf{0.9917\,$\pm$\,0.01} & \textbf{0.1583\,$\pm$\,0.17} & 0.3868\,$\pm$\,0.07 & 0.3632\,$\pm$\,0.07 & 0.9394\,$\pm$\,0.06 & 0.6132\,$\pm$\,0.07 & 0.1343\,$\pm$\,0.05 & 0.1049\,$\pm$\,0.03 & 0.8005\,$\pm$\,0.14 & 0.8657\,$\pm$\,0.05 \\
50 & \textbf{0.9461\,$\pm$\,0.12} & \textbf{0.9402\,$\pm$\,0.12} & \textbf{0.9940\,$\pm$\,0.01} & \textbf{0.0539\,$\pm$\,0.12} & 0.5049\,$\pm$\,0.10 & 0.4956\,$\pm$\,0.10 & 0.9809\,$\pm$\,0.03 & 0.4951\,$\pm$\,0.10 & 0.1466\,$\pm$\,0.05 & 0.1103\,$\pm$\,0.03 & 0.7816\,$\pm$\,0.15 & 0.8534\,$\pm$\,0.05 \\
\bottomrule
\end{tabular}}
\end{table*}

Before any attack ($k{=}0$), ensemble~A correctly identifies 87.5\% of malware samples. This is the baseline that all three methods must overcome. The CVAE consistently outperforms both baselines at every $k$ tested. Even a small $k{=}5$ already reduces $\mathrm{Recall}_6$ from 0.87 to 0.75. By $k{=}15$, a majority of malware samples evade detection ($\mathrm{Recall}_6 = 0.44$). At $k{=}20$, approximately 7~out of~10 samples evade ($\mathrm{Recall}_6 = 0.30$). By $k{=}50$, the CVAE achieves near-total evasion ($\mathrm{Recall}_6 = 0.05$). For comparison, with $k{=}20$ injected imports MostPopular reaches only $\mathrm{Recall}_6 = 0.69$, while Random leaves $\mathrm{Recall}_6$ essentially unchanged at 0.87.

The gap between CVAE and MostPopular is especially significant given that both methods use the same target class and inject the same number of calls. The difference comes from how features are chosen. MostPopular always selects the same $k$ most frequent APIs of the target class regardless of the input, whereas the CVAE selects different APIs for each malware sample based on which calls are already present and which additions would be most effective. The Random baseline shows that arbitrary additions have essentially no effect on detection. This confirms that targeted selection of API calls, not the number of additions, drives evasion success. The CTS of MostPopular grows with $k$, from 0.85 at $k{=}5$ to 0.98 at $k{=}50$. This shows that even a frequency-based strategy becomes more precisely targeted when more imports are 
injected, although its overall evasion rate stays substantially below that of the CVAE.

Importantly, evasion is not merely untargeted but precisely targeted. The CVAE's UER and TSR remain closely correlated across all $k$ tested (e.g., 0.95 vs.\ 0.94 at $k{=}50$), and CTS reaches ${\approx}0.99$ for $k \geq 20$. This means that when the CVAE evades detection, it almost always falls into a specific benign class selected by the ensemble~B, rather than scattering across random benign categories. This highlights a practical vulnerability in API-based malware classifiers.

\begin{figure*}[!t]
\centering
\includegraphics[width=0.84
\textwidth]{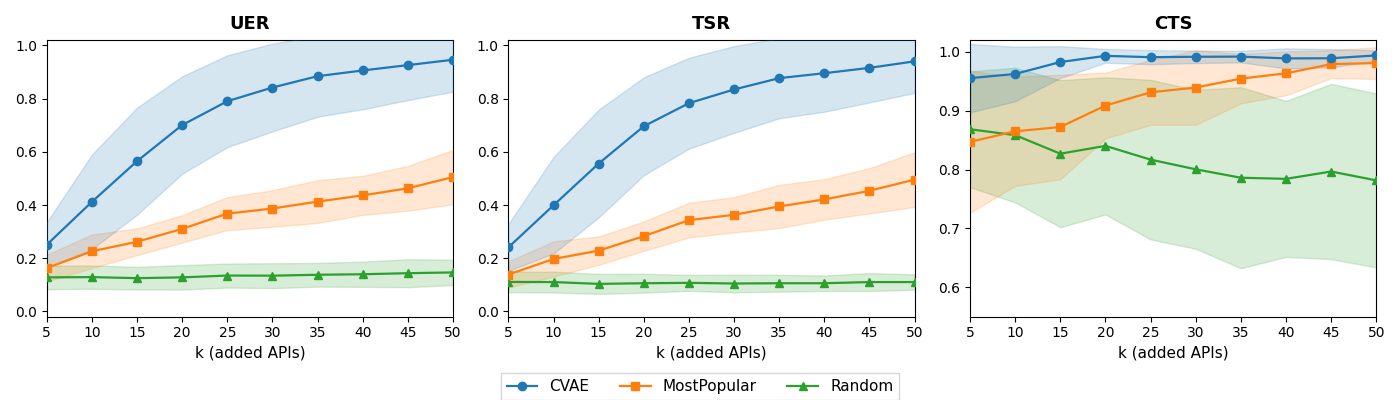}
\caption{Evasion performance (UER, TSR, CTS) across different values of $k$  (number of injected API calls). Lines indicate the mean over 20 runs, and shaded bands span the interquartile range (IQR).}
\label{fig:evasion_curves}
\end{figure*}

While Table~\ref{tab:comparison} reports mean values, Fig.~\ref{fig:evasion_curves} shows the spread of each metric across the 20 runs. The CVAE lies above both baselines on UER and TSR at every $k$, with the gap widening as $k$ increases. The CTS bands of CVAE and MostPopular overlap at higher $k$, but this convergence is misleading. MostPopular reaches CTS $\approx 0.98$ at $k{=}50$, but its UER is still only $\approx 0.50$ at that point, so just half of the malware samples evade detection, and of those, 98\% land in the target class. By contrast, the CVAE attains comparable CTS already at $k{=}5$ (0.96 vs.\ 0.85) and reaches near-perfect targeting from $k \geq 20$.

\begin{figure}[!t]
\centering
\includegraphics[width=0.83\columnwidth]{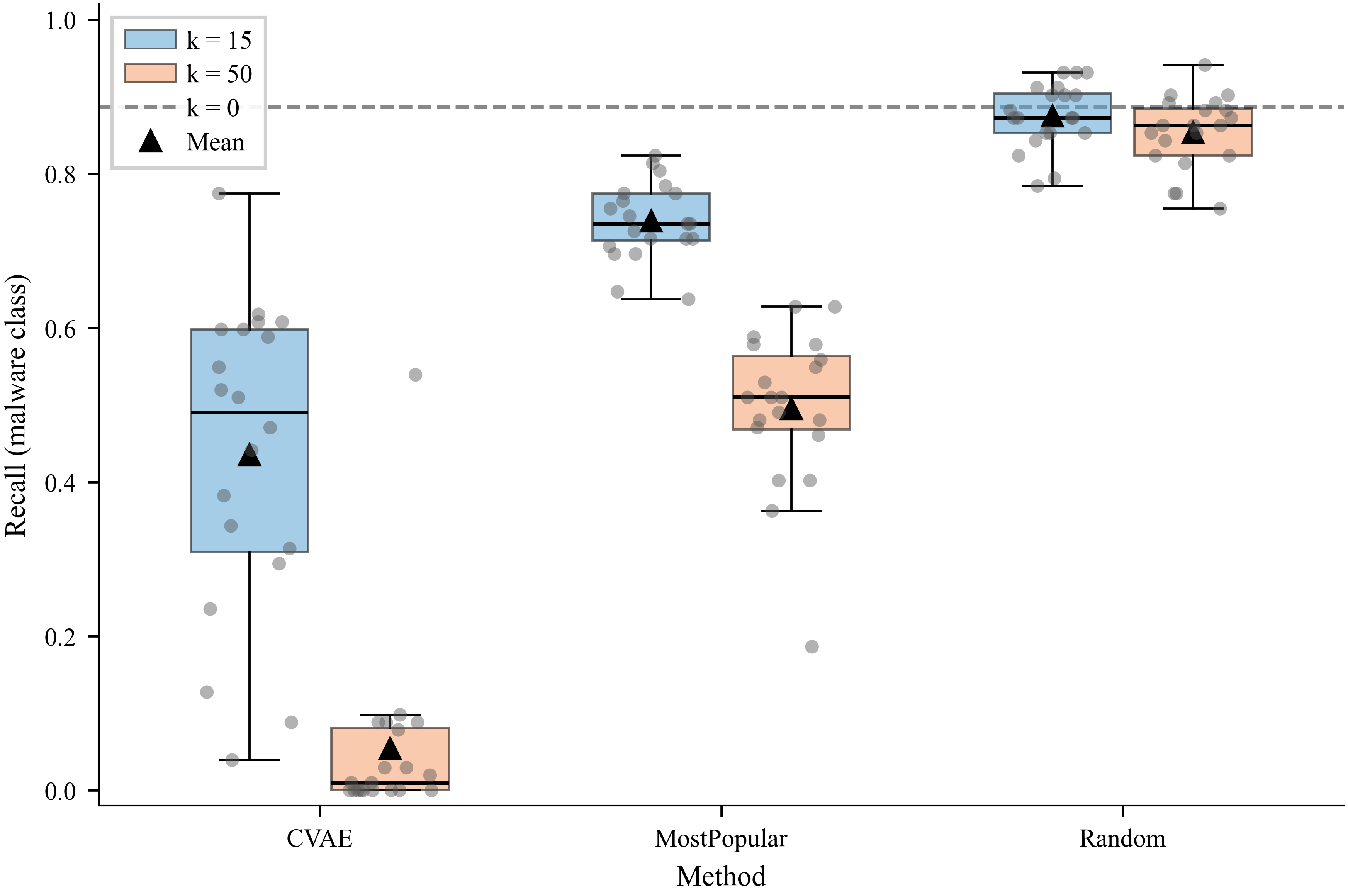}
\caption{Distribution of $\mathrm{Recall}_6$ at $k{=}15$ and $k{=}50$ across 20 runs for all three methods.}
\label{fig:recall_curve}
\end{figure}

Fig.~\ref{fig:recall_curve} shows the per-run distribution of $\mathrm{Recall}_6$ at $k{=}15$ and $k{=}50$. At $k{=}15$, the CVAE's box lies entirely below MostPopular's, with no overlap between the two methods. At $k{=}50$, the CVAE's distribution concentrates tightly near zero, while MostPopular and Random remain well above. Even the worst-performing CVAE run at $k{=}50$ falls below the MostPopular mean, showing that the advantage holds not only on average but also in the tail of the distribution.

In addition to the numerical metrics, Fig.~\ref{fig:tsne_panels} provides a geometric view of what evasion looks like in feature space. In the raw feature space (panel~a), classes overlap substantially, making classification difficult. After encoder training (panel~b), the learned embeddings form well-separated clusters for all 6 classes, demonstrating that the encoder captures meaningful class structure. Panel~(c) shows the embedding space after adversarial modification at $k{=}50$, most malware samples have migrated toward their assigned benign target clusters, confirming that the CVAE shifts the malware representation deep into the target class region of the embedding space.

\subsection{Validation on Real PE Files via VirusTotal}
\label{sec:virustotal}

To verify that the attack transfers beyond our experimental classifier, we modified ten malware PE files from dataset by injecting the CVAE-selected API imports into the IAT and submitted both the original and modified versions to VirusTotal. Malware functionality is preserved by appending two new sections. A new data section holds a rebuilt import table containing both the original and the CVAE generated imports. A new code section holds one trampoline per original import. Each original IAT entry is redirected to its trampoline, which jumps indirectly through the corresponding slot in the rebuilt IAT. The Windows loader resolves that rebuilt IAT at load time, so every original call site reaches its unmodified target function allowing PE file to run properly.

Table~\ref{tab:virustotal} shows results at $k{=}50$. The average number of engines flagging a sample as malicious dropped from 48 to 22, a mean reduction of 54.5\%, with per-sample reductions ranging from 31.4\% to 75.0\%. Since the CVAE was not trained against any of these engines, this constitutes a black-box transfer test. Commercial engines employ diverse techniques beyond static API analysis, including signatures, heuristics, and behavioural analysis, yet the observed reduction confirms that API import patterns contribute to real-world static detection decisions.

\begin{table}[!t]
\centering
\caption{VirusTotal detection counts before and after API import injection ($k{=}50$, out of ${\sim}$71 engines).}
\label{tab:virustotal}
\small
\begin{tabular}{lcccc}
\toprule
\textbf{Sample} & \textbf{Before} & \textbf{After} & \textbf{$\Delta$} & \textbf{Reduction} \\
\midrule
Sample 1  & 53 & 28 & $-$25 & 47.2\% \\
Sample 2  & 56 & 33 & $-$23 & 41.1\% \\
Sample 3  & 52 & 20 & $-$32 & 61.5\% \\
Sample 4  & 29 & 14 & $-$15 & 51.7\% \\
Sample 5  & 46 & 19 & $-$27 & 58.7\% \\
Sample 6  & 51 & 35 & $-$16 & 31.4\% \\
Sample 7  & 46 & 17 & $-$29 & 63.0\% \\
Sample 8  & 51 & 23 & $-$28 & 54.9\% \\
Sample 9  & 48 & 12 & $-$36 & 75.0\% \\
Sample 10 & 55 & 22 & $-$33 & 60.0\% \\
\midrule
\textbf{Mean} & \textbf{48.7} & \textbf{22.3} & $\bm{-26.4}$ & \textbf{54.5\%} \\
\bottomrule
\end{tabular}
\end{table}

\section{Limitations and Future Work}
\label{sec:future}

While the proposed framework demonstrates that targeted API import injection is a concrete and exploitable vulnerability in static malware detection, several limitations constrain the scope of this study. The framework operates exclusively on static IAT entries, so samples that evade static classification may still be caught by dynamic monitoring of system calls, network behaviour, or process activity. We assume a grey-box threat model with access to the feature representation and training data but not the detector's internal parameters; effectiveness under a query-only black-box setting remains open. Finally, the evaluation is conducted on a single six-class Windows PE dataset, leaving generalisation to other feature representations, larger class taxonomies, or different malware families to future work.

Future directions include extending the framework to dynamic behavioural features such as system call sequences to test whether targeted evasion remains feasible against detectors combining static and dynamic analysis, applying the same framework defensively to generate adversarial samples for adversarial training, and investigating whether additive injection generalises to other static features such as raw bytes or PE header fields. 

This research is intended to expose weaknesses in existing malware detection systems so that they can be addressed by the security community. No adversarial samples were deployed against production systems or distributed outside the research team.

\section{Conclusion}
\label{sec:conclusions}

This paper demonstrated that API-based malware classifiers are vulnerable not only to evasion but to targeted evasion. An attacker can systematically force misclassifications into a specific chosen benign software category by adding a small number of Win32 API imports, without removing existing imports or accessing the detector's internal parameters. The attack is implemented by a CVAE whose decoder can only inject new imports, never remove existing ones, preserving malware functionality by design. The CVAE is guided by an ensemble that selects each sample's most plausible benign target and trained against a knowledge-distilled differentiable proxy of the detector. To support this study, we constructed and released a six-class dataset of Win32 API import vectors from 3,799 Windows executables.

The proposed API import injection attack is highly effective against the ensemble detector, which correctly identifies 87.5\% of malware samples before modification. Adding just 20 API imports reduces malware recall to 0.30, with 70\% of malware samples misclassified as the intended target category and a conditional target success of 99\%. At 50 added imports, recall drops to 5\%. The CVAE substantially outperforms the MostPopular and Random baselines at every tested $k$, demonstrating that which API calls are injected matters more than how many. Validation on real PE files via VirusTotal confirms that the attack transfers to commercial static detection engines, averaging a 54.5\% reduction in flagging engines.

These findings expose a concrete and exploitable weakness in static, API-based detection. Detectors can be deceived with as few as 5--20 injected API imports, without modifying existing code or accessing the detector's internal parameters. Moreover, the misclassification can be directed toward a chosen benign category with high precision, turning evasion from a generic ``not malware'' outcome into a directed disguise. Two defensive directions follow. First, static API analysis should be combined with dynamic behavioural monitoring to reduce reliance on import patterns. Second, detector training should incorporate additive adversarial samples to improve robustness against import injection.

\section*{Acknowledgments}
This project has received funding from the Research Council of Lithuania (LMTLT), agreement No S-MIP-24-116.

\bibliographystyle{IEEEtran}
\bibliography{bibliography}

\vspace{-30pt} 
\begin{IEEEbiography}[{\includegraphics[width=1in,height=1in,clip,keepaspectratio]{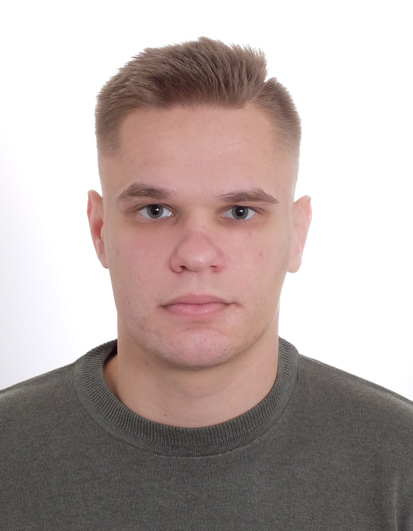}}]{Juozas Dautartas} is a Ph.D. student in Informatics Engineering at Vilnius University, Institute of Data Science and Digital Technologies. His research focuses on adversarial machine learning, malware analysis, and ML-based cybersecurity. He is the principal initiator and lead contributor of the WinAPI-AdvMal dataset, the first multi-category adversarial malware dataset for Windows PE files. He has co-authored several publications at IEEE conferences, receiving the Outstanding Oral Presentation Award at AAIML 2026 in Tokyo, Japan.
\end{IEEEbiography}
\vspace{-33pt} 
\begin{IEEEbiography} [{\includegraphics[width=1in,height=1.25in,clip,keepaspectratio]{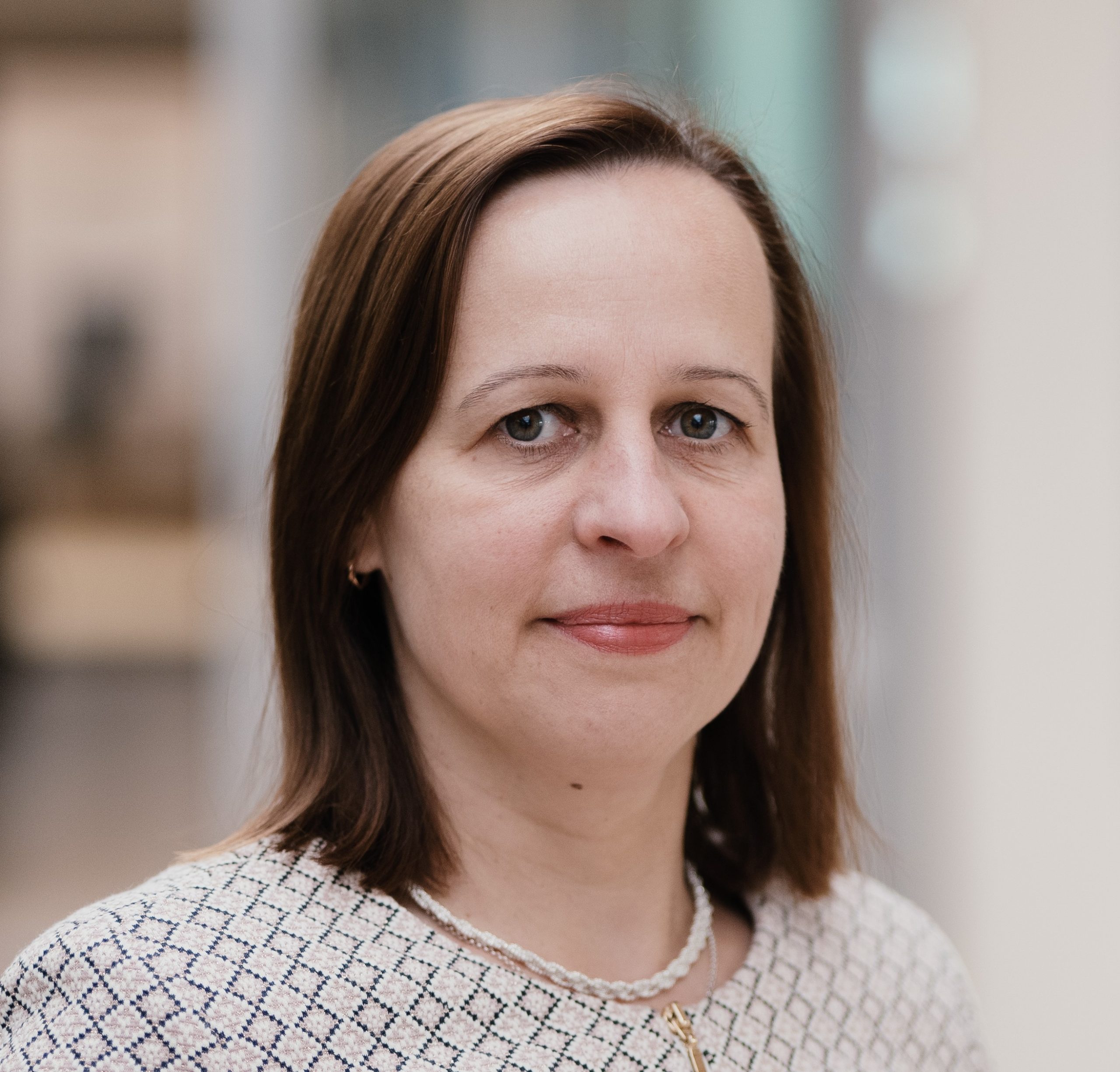}}]{Olga Kurasova} is a Professor and Principal Researcher  at Vilnius University, Institute of Data Science and Digital Technologies. Her research areas include artificial intelligence, machine learning, and multidimensional data visualization. She has published more than 120 scientific articles and has supervised numerous doctoral students. She is a Full Member of the Lithuanian Academy of Sciences, a Senior Member of IEEE, and a recipient of the Lithuanian Science Prize 2021 and the Vilnius University Rector's Science Prize~2024.
\end{IEEEbiography}
\vspace{-33pt} 
\begin{IEEEbiography}[{\includegraphics[width=1in,height=1in,clip,keepaspectratio]{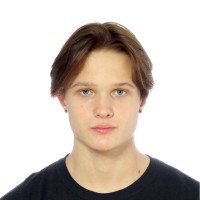}}]{Juozapas Rokas Čypas} is a Research Assistant at Vilnius University, Institute of Data Science and Digital Technologies, where he works at the intersection of cybersecurity and machine learning. His research interests include software engineering and machine learning, with a particular focus on malware analysis, adversarial machine learning and offensive AI. He has contributed to several peer-reviewed publications and research projects in the field of cybersecurity.
\end{IEEEbiography}
\vspace{-33pt} 
\begin{IEEEbiography}[{\includegraphics[width=1in,height=1in,clip,keepaspectratio]{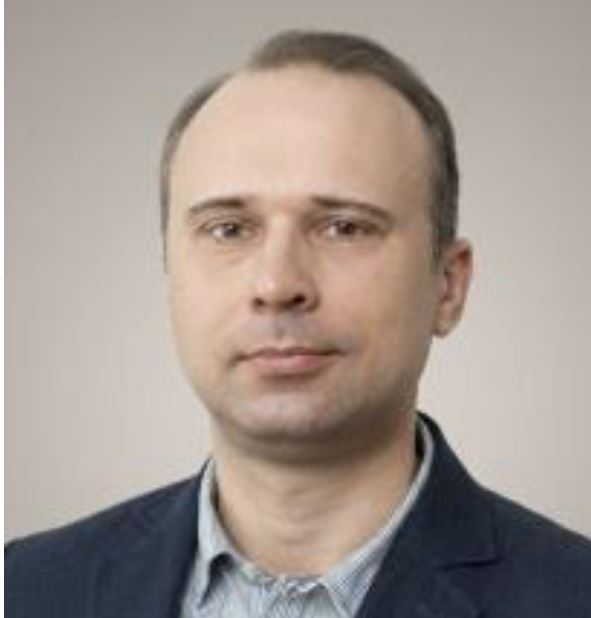}}]{Viktor Medvedev} is a Senior Researcher and Associate Professor at Vilnius University, Institute of Data Science and Digital Technologies. His research interests include artificial intelligence, deep learning, cybersecurity, adversarial machine learning, data analysis, and multidimensional data visualisation. He has authored over 50 peer-reviewed publications and has supervised doctoral students in cybersecurity and offensive AI.
\end{IEEEbiography}

\vfill

\end{document}